\documentclass[preprint,12pt]{elsarticle}

\RequirePackage{luatex85}
\usepackage{graphicx}
\usepackage{amssymb}
\usepackage{amsmath}
\usepackage{bm}
\usepackage{lineno}
\usepackage{float}
\usepackage{multicol}
\usepackage{multirow}
\usepackage{tabularx}
\usepackage{booktabs}

\journal{Physical Review Fluids}

\begin{document}
\begin{frontmatter}

\title{On correcting the eddy-viscosity models in RANS simulations for turbulent flows and scalar transport around obstacles}

\author{Zengrong Hao, Catherine Gorl\'e, David S. Ching, John K. Eaton}

\address{Stanford University, Stanford, CA 94305, USA}

\begin{abstract}
In Reynolds-averaged-Navier-Stokes (RANS) simulations for turbulent scalar transport, it is common that using an eddy-viscosity (EV) model to close the Reynolds stress yields reasonable mean flow predictions but large errors in scalar transfer results regardless of scalar flux model inadequacies. This failure mode of EV models is generally related to the fact that the transport of momentum and scalar depends on different Reynolds stress components. The present work addresses two common issues relevant to such failures in the scenarios of turbulent scalar transport around obstacles. The first issue is the general overprediction of scalar transfer near the upwind surfaces, which is primarily attributed to the absence of wall-blocking mechanism in conventional EV models. We accordingly propose a Shear-Preserving-Wall-Blocking (SPWB) method to analytically correct the overpredicted wall-normal stress under the realizability constraint. The second issue is the general underprediction of scalar transfer in the downstream large separation regions, which is essentially attributed to the presence of vortex shedding invalidating the scaling ground in conventional EV models' dissipation closures. We accordingly generalize the recently proposed Double-Scale Double-Linear-EV (DSDL) model to scalar transport predictions. Consequently, a combined model, SPWB-DSDL, is developed. The model is then applied to two test cases, of which the first features a bluff obstacle with an upstream impingement flow and a downstream two-dimensional separation and the second features a streamlined obstacle with an upstream concave surface flow and a downstream three-dimensional separation. In the two cases, the SPWB-DSDL model is capable of simultaneously yielding reasonable results of mean flow field, turbulence energy and stress, and scalar transfer in both upstream and downstream regions, thus demonstrating significant improvement upon a conventional EV model with or without stress shape perturbations.

\end{abstract}


\end{frontmatter}


\section{Introduction}
\label{Sec:Introduction}

Turbulent flows with scalar (e.g. temperature, concentration, and humidity) transport are widespread in urban wind engineering, such as pollutant dispersion, natural ventilation, and urban heat island mitigation. Confined by and interacting with the complex urban landscape, the scalar transport can exhibit diverse regimes in different spatial regions. This diversity poses a fundamental challenge to the Reynolds-averaged-Navier-Stokes (RANS) modeling for turbulence, especially for those models based on the simple eddy-viscosity (EV) assumption that is dominantly adopted in engineering applications. While an established EV model can be capable of predicting a satisfactory mean flow field, it may still lead to unacceptable results of some scalar-related quantities of interest (QoIs) regardless of scalar flux model inadequacies. For this failure mode of EV models, to identify the specific failure mechanisms responsible is essential for us to understand the real-world turbulence physics and in turn improve the models. In this paper, we consider two common issues relevant to such failures in urban wind engineering.

The first issue is on the scalar transport on the upwind surface of a structure. We know that in some flows with relatively high free-stream turbulence intensity (typically \(>5\%\)), many conventional EV models predict nonphysically large production of turbulence kinetic energy \(k\) around the stagnation points \cite{Strahle85,Launder93}, named `stagnation point anomaly'. This defect may not seriously affect the mean flow predictions, but can directly yield significant overpredictions (up to \(300\%\)) of scalar transfer rates (e.g.~\cite{Craft93,Ashforth96,Behnia98,Parneix99,Park01,Sagot08,Zu09,Sharif13,Szczepanik04,Li16,Hao20b}). To alleviate this problem of EV models, one approach is to design some \textit{ad hoc} limiters to the eddy viscosity \(\nu_t\) (e.g.~\cite{Kato93,Menter94,Durbin96,Park97}). This approach, however, does not essentially address a more elemental reason for the overpredictions: Without considering the wall-blocking effect, an EV model tends to overestimate the wall-normal component of Reynolds stress, symbolically \(\overline{v^{\prime2}}\), close to walls, especially in the region with strong mean flow contraction in wall-normal direction (as indicated in \cite{Craft93,Behnia98,Sharif13}). Another approach is to directly model the wall-blocking effect on the transport of \(\overline{v^{\prime2}}\). This approach features the notable `elliptic relaxation models' \cite{Durbin91,Durbin95,Parneix98,Behnia98}. However, in practice, the elliptic equation is potentially a major burden to a CFD solver in terms of convergence and efficiency, especially in large-scale applications. 

The second issue is on the scalar transport around the large separation and recirculation regions behind a structure. As reported in \cite{Parente11a,Garnier12,Zeoli18,Hao20a}, steady RANS simulations with EV models can significantly underpredict (by up to 90\%) \(k\) around these regions, thereby nonphysically suppressing the possible scalar transport there (e.g.~\cite{Hao20b}). As analyzed in \citet{Hao21}, these underpredictions can be essentially attributed to the failure of EV models to distinguish those fluctuations induced by vortex shedding, a typical form of coherent structures (CS) that are semi-deterministic and quasi-periodic, from the background stochastic turbulence (ST). They accordingly proposed a three-equation Double-Scale Double-Linear-eddy-viscosity (DSDL) model to separately characterize the dissipation behaviors of CS and ST within the steady RANS framework. The DSDL model showed significant improvement of \(k\) predictions around the large separation and recirculation regions behind several types of bluff bodies. These favorable results imply the potential of DSDL model to also improve the scalar transport prediction between the mainstream and the large recirculation region behind an urban structure.

In this paper, the above two issues on the EV model failures in turbulent scalar transport predictions are addressed based on flow physics. For the first issue, we propose a method quantifying the wall-blocking effect by analytically correcting the wall-normal stress component under realizability constraint while preserving the shear stress component; this method is denoted by `Shear-Preserving-Wall-Blocking' (SPWB). For the second issue, we directly generalize the recently proposed DSDL model \cite{Hao21} to scalar transport predictions, with a slight reformulation of the energy transfer rate function. An integrated methodology `SPWB-DSDL' is thus developed. In addition, for the sake of comparison, a previously developed method for correcting the EV models, termed `Reynolds-Stress-Shape-Perturbation' (RSSP) \cite{Emory13,Gorle12}, is also considered (separately from SPWB and DSDL). Subsequently, with the popular \(k\)-\(\omega\) SST model \cite{Menter09} as baseline, these model corrections are applied to two test cases. The first case is the forced heat convection in a pin-fin array, which features bluff obstacles with upstream impingement flows and downstream quasi-two-dimensional separations; the second case is the scalar dispersion around a skewed bump mounted on a wall, which features a streamlined obstacle with an upstream concave surface flow and a downstream complicated three-dimensional separation.

For the closure of scalar flux, we use the generalized-gradient-diffusion-hypothesis (GGDH) model \cite{Daly70} for all simulations of this paper. Compared to the more widely used standard-gradient-diffusion-hypothesis (SGDH) model \cite{Batchelor49}, GGDH is generally recognized as more reasonably reflecting the influence of hydrodynamic turbulence anisotropy on the scalar diffusivity anisotropy. Furthermore, we boldly assume that the GGDH model inadequacies are qualitatively less influential than the conventional EV model inadequacies (see the comparison of different scalar flux models in \cite{Hao20b}) and thus not discussed at this stage without influencing the conclusions.

The remainder of this paper is organized as follows. Section \ref{Sec:Baseline} includes the basic RANS equations and the baseline model adopted in this paper. Section \ref{Sec:Methods} introduces the methodologies of SPWB and DSDL, together with a brief review of the RSSP for comparison. Section \ref{Sec:Applications} applies these methodologies in the two test cases, and analyzes the results through comparisons with the high-fidelity reference databases. Section \ref{Sec:Conclusions} summarizes the methodologies and results of this paper.

\section{RANS equations with the baseline model}
\label{Sec:Baseline}

Consider an incompressible flow (with homogeneous density and constant viscosity \(\nu\)) carrying a passive, source-free scalar (with constant diffusion coefficient \(\kappa\)). The RANS equations for mean velocity \(\textit{\textbf{U}}\), mean kinematic pressure \(P\) and mean scalar \(\mathit{\Theta}\) are
\begin{subequations}
\label{eq:basicRANS}
\begin{align}
    & \nabla\cdot\textit{\textbf{U}} = 0\,, \\
    & \bar{D}_t\textit{\textbf{U}} = -\nabla P+\nabla\cdot\left(\,\nu\nabla\textit{\textbf{U}}-\textsf{\textbf{R}}\,\right), \\
    & \bar{D}_t\mathit{\Theta} = \nabla\cdot\left(\,\kappa\nabla\mathit{\Theta}-\textit{\textbf{q}}\,\right),
\end{align}
\end{subequations}
where \(\bar{D}_t\equiv\partial_t+\textit{\textbf{U}}\cdot\nabla\) is material derivative, \(\textsf{\textbf{R}}\equiv\overline{\textit{\textbf{u}}^\prime\textit{\textbf{u}}^\prime}\) is the tensor of Reynolds stress, and \(\textit{\textbf{q}}\equiv\overline{\textit{\textbf{u}}^\prime\theta^\prime}\) is the vector of scalar flux.

A popular EV model, \(k\)-\(\omega\) SST model \cite{Menter09}, is used as the baseline model to close \(\textsf{\textbf{R}}\):
\begin{subequations}
\label{eq:EV}
\begin{align}
    & \textsf{\textbf{R}} = 2k\,\textsf{\textbf{I}}/3 - 2\nu_t\,\textsf{\textbf{S}}\,, \quad \textsf{\textbf{S}}\equiv (\nabla\textit{\textbf{U}}+\left(\nabla\textit{\textbf{U}}\right)^\mathrm{T})/2\,, \\
    & \nu_t = k\,/\,\mathrm{max}(\omega,SF_2/a_1)\,, \quad S\equiv\sqrt{2\,\textsf{\textbf{S}}:\textsf{\textbf{S}}}
\end{align}
\end{subequations}
with
\begin{subequations}
\label{eq:SSTmodel}
\begin{align}
    & \bar{D}_tk = \mathcal{P}_\mathrm{lim} - \varepsilon + \nabla\cdot\left[\,(\nu+\sigma_k\nu_t)\nabla k\,\right], \\
    & \bar{D}_t\omega = \left[\,\alpha\,\mathcal{P}_\mathrm{lim} - (\beta/C_\mu)\,\varepsilon\,\right]/\nu_t + \nabla\cdot\left[\,(\nu+\sigma_\omega\nu_t)\nabla\omega\,\right] + C\!D\,, \\
    & \mathcal{P}_\mathrm{lim} = \mathrm{min}\{\mathcal{P},\, 10\,\varepsilon\},
\end{align}
\end{subequations}
where \(\textsf{\textbf{I}}\) is the identity tensor of order two; \(\mathcal{P}\equiv-\textsf{\textbf{R}}:\textsf{\textbf{S}}\) and \(\varepsilon \equiv C_\mu\omega k\) are respectively the production and the dissipation rates of \(k\); the constant \(C_\mu=0.09\); and the details of term \(C\!D\), blending function \(F_2\), and coefficients \(a_1\), \(\sigma_k\), \(\sigma_\omega\), \(\alpha\) and \(\beta\) can be found in \cite{Menter09}. This baseline EV model will be taken for example to expound those correction methodologies in Section \ref{Sec:Methods} and applied in Section \ref{Sec:Applications}.

The scalar flux \(\textit{\textbf{q}}\) is closed by the algebraic GGDH model \cite{Daly70} (written in terms of \(\nu_t\)):
\begin{equation}
\label{eq:GGDHmodel}
    \textit{\textbf{q}} = -\left(c_\theta\,\nu_t/C_\mu\right)\left(\textsf{\textbf{R}}/k\right)\cdot\nabla\mathit{\Theta}\,, \quad c_\theta=0.3\,.
\end{equation}

\section{Methodologies of EV model corrections}
\label{Sec:Methods}

\subsection{Shear-Preserving Wall-Blocking (SPWB) correction}
\label{Subsec:SPWB}

\subsubsection{Analysis of wall-blocking effect on scalar transport}
\label{Subsubsec:SPWBmotive}

Before discussing the correction methodology, let us first qualitatively analyze how an EV model without wall-blocking corrections ultimately results in the errors of scalar transport predictions.

Consider a shear-dominated, quasi-planar mean flow with shear rate \(\partial_y U>0\) and spanwise strain rate \(\partial_z W\simeq0\); impose a passive scalar field with mean gradient \(\partial_y\mathit{\Theta}>0\). Historically, the primary concern in an EV model development is to predict the shear component of Reynolds stress for this type of mean flows via \(\overline{u^\prime v^\prime}=-\nu_t\,\partial_yU\). The objective can be formally achieved by adjusting the nominal eddy viscosity \(\nu_t\). However, this adjusted \(\nu_t\) value, as we know, usually leads to incorrect normal stress components through
\begin{subequations}
\label{eq:EVnormal}
    \begin{align}
        & \overline{u^{\prime2}}=2k/3-2\nu_t\,\partial_xU\,, \\
        & \overline{v^{\prime2}}=2k/3-2\nu_t\,\partial_yV\,.
    \end{align}
\end{subequations}
Subsequently, the error of Eq.~(\ref{eq:EVnormal}) can influence the transverse (i.e.~in \(y\) direction) scalar transport typically via the following two mechanisms.

The first mechanism is through the \(k\) production
\begin{equation}
\label{eq:kProd}
    \mathcal{P}\simeq -\overline{u^\prime v^\prime}\,\partial_yU+\mathcal{P}_\mathrm{imb}
\end{equation}
where
\begin{equation}
\label{eq:Pimb}
    \mathcal{P}_\mathrm{imb}\equiv-\overline{u^{\prime2}}\,\partial_xU-\overline{v^{\prime2}}\,\partial_yV\simeq\left(\overline{u^{\prime2}}-\overline{v^{\prime2}}\right)\partial_yV
\end{equation}
is the extra production induced by the normal stress imbalance. The EV relation (\ref{eq:EVnormal}) yields \(\mathcal{P}_\mathrm{imb}=4\nu_t\left(\partial_yV\right)^2\), which is always non-negative. The problem of this result is \textit{especially evident in some near-wall flows with strong mean contraction in wall-normal direction (i.e.~\(\partial_yV<0\))}, where in reality \(\overline{v^{\prime2}}\) is considerably smaller than \(\overline{u^{\prime2}}\) due to the wall-blocking effect and thus \(\mathcal{P}_\mathrm{imb}\) should be negative. Therefore, in such flows, an EV model without special treatments can substantially overestimate \(\mathcal{P}_\mathrm{imb}\) and thus \(k\) (e.g.~the `stagnation point anomaly' can be regarded as an extreme example), and in consequence overpredict the scalar transport there.

The second mechanism is more direct. In physics, the transverse scalar transport (measured by scalar flux \(\overline{v^\prime\theta^\prime}\)) directly depends on the transverse velocity fluctuation (\(\overline{v^{\prime2}}\)) but not directly depends on the streamwise fluctuation (\(\overline{u^{\prime2}}\)) or other cross correlations (\(\overline{u^\prime v^\prime}\)); this fact is also reflected by the GGDH model (\ref{eq:GGDHmodel}), which indicates \(\overline{v^\prime\theta^\prime}\propto\overline{v^{\prime2}}\). Therefore, an overpredicted \(\overline{v^{\prime2}}\) immediately results in the overprediction of (magnitude of) \(\overline{v^\prime\theta^\prime}\). As in the first mechanism, this overprediction is \textit{especially severe in the near-wall flows with \(\partial_yV<0\)}, where the real \(\overline{v^{\prime2}}\) is subjected to the wall-blocking effect and thus can be significantly smaller than the value given by Eq.~(\ref{eq:EVnormal}b).

The above qualitative analysis indicates that the ignorance of wall-blocking effect is plausibly the primary reason for the EV models' overprediction of scalar transport on the upwind surface of a structure, on which the boundary layer is usually subjected to a strong wall-normal contraction. Therefore, a general wall-blocking correction to EV models is essential to address this overprediction.

\subsubsection{Stress components in local shear-normal-span coordinate system}
\label{Subsubsec:tbnCoordSys}

Compared to the global coordinate system \(x\)-\(y\)-\(z\), it is more natural to discuss the wall-blocking effect in a local coordinate system that depends on the local wall geometry. Here we define a local `shear-normal-span' system and express the Reynolds stress components in it before implementing corrections in the remainder of \S\ref{Subsec:SPWB}.

Consider a Reynolds stress tensor \(\textsf{\textbf{R}}\) at a point. The local wall-normal unit vector \(\textbf{\textit{n}}\) can be calculated via \(\textbf{\textit{n}} = \nabla d_w\) where \(d_w\) is the (field of) wall distance. Then the shear stress vector acting on the plane perpendicular to \(\textbf{\textit{n}}\) is
\begin{equation}
\label{eq:RSsh}
    \bm{\tau} = (\textsf{\textbf{I}}- \textbf{\textit{n}}\textbf{\textit{n}})\cdot(\textbf{\textit{n}}\cdot\textsf{\textbf{R}}).
\end{equation}
We define the local shear unit vector as \(\textit{\textbf{t}}=\bm{\tau}/\tau\) where
\begin{equation}
\label{eq:tauDef}
    \tau\equiv-\parallel\bm{\tau}\parallel
\end{equation}
and the local spanwise unit vector as \(\textit{\textbf{b}}=\textit{\textbf{t}}\times\textit{\textbf{n}}\). The vectors \((\textit{\textbf{t}},\textit{\textbf{n}}, \textit{\textbf{b}})\) form the bases of a local orthogonal coordinate system (named `\(t\)-\(n\)-\(b\) system') with its Jacobian tensor
\begin{equation}
\label{eq:tensorQ}
    \textsf{\textbf{Q}} = \textit{\textbf{t}}\textit{\textbf{e}}_x + \textit{\textbf{n}}\textit{\textbf{e}}_y + \textit{\textbf{b}}\textit{\textbf{e}}_z
\end{equation}
where \((\textit{\textbf{e}}_x,\textit{\textbf{e}}_y,\textit{\textbf{e}}_z)\) are the bases of the global \(x\)-\(y\)-\(z\) system. Thereby the components of \(\textsf{\textbf{R}}\) in the local \(t\)-\(n\)-\(b\) system are contained in the matrix of the transformed tensor
\begin{equation}
\label{eq:transform}
    \widetilde{\textsf{\textbf{R}}} = \textsf{\textbf{Q}}^\mathrm{T}\cdot\textsf{\textbf{R}}\cdot\textsf{\textbf{Q}}\,.
\end{equation}
The matrix of \(\widetilde{\textsf{\textbf{R}}}\), which we denote by \([\,\widetilde{\textsf{\textbf{R}}}\,]\), has the form
\begin{equation}
\label{eq:matrixRtilde}
    [\,\widetilde{\textsf{\textbf{R}}}\,] =
    \left[
    \begin{array}{c c c}
        \sigma_t & \tau & \tau_b \\
        \tau & \sigma_n & 0 \\
        \tau_b & 0 & \sigma_b
    \end{array}
    \right]
\end{equation}
in which \(\sigma_t\), \(\sigma_n\), and \(\sigma_b\) are the normal stress components respectively in the \(t\), \(n\), and \(b\) directions; \(\tau\) (defined by (\ref{eq:tauDef})) is the \(t\)-direction shear stress acting on the \(t\)-\(b\) plane; and \(\tau_b\) is the \(t\)-direction shear stress acting on the \(t\)-\(n\) plane, which can be nonzero if the mean flow is inhomogeneous in the \(b\) direction.

Fig.~\ref{fig:SPWBprinciple}(a) illustrates the two-dimensional stress state of \([\,\widetilde{\textsf{\textbf{R}}}\,]\) on the \(t\)-\(n\) plane. This state can be graphically represented by a Mohr's circle's diameter line \(\overline{AB}\) with the endpoint coordinates \(A(\sigma_t,\tau)\) and \(B(\sigma_n,-\tau)\) and the centre coordinates \(C((\sigma_t+\sigma_n)/2,0)\) (Fig.~\ref{fig:SPWBprinciple}(b)). Denote the intersections of the circle with the horizontal axis by \((\sigma_\mathrm{min},0)\) and \((\sigma_\mathrm{max},0)\). Then the realizability constraint (i.e.~the semi-positive-definiteness of a Reynolds stress) is equivalent to
\begin{subequations}
\label{eq:Rlz}
    \begin{align}
        & \sigma_\mathrm{min}\geq0\,, \\
        & (\sigma_t+\sigma_n)/2\leq k\,, \\
        & \tau_b^2 \leq \sigma_t\,\sigma_b\,.
    \end{align}
\end{subequations}
The remainder of \S\ref{Subsec:SPWB} will discuss the wall-blocking correction to the stress matrix \([\,\widetilde{\textsf{\textbf{R}}}\,]\) by means of the above Mohr's circle representation.

\subsubsection{Correcting the wall-normal stress while preserving the shear stress}
\label{Subsubsec:SPWBmethod}

Let us suppose that the wall-blocking correction methodology to be proposed have the following five properties:

\renewcommand{\theenumi}{(\Roman{enumi})}
\begin{enumerate}

\item \textit{The realizability constraint (\ref{eq:Rlz}) is met.}

\item \textit{The shear stress component \(\tau\) is preserved.} The reasoning behind this argument is that the predictions of the shear stresses in wall-bounded flows are usually the primary concern in an EV model development (as discussed at the beginning of \S\ref{Subsubsec:SPWBmotive}), and thus have relatively higher credibility than the model's predictions of other components.

\item \textit{The turbulence kinetic energy \(k=(\sigma_t+\sigma_n+\sigma_b)/2\) is preserved.} The reasoning behind it can be perceived through the following fact: the wall-blocking effect on the Reynolds stress transport is via the harmonic pressure-strain rate correlation, which acts to redistribute Reynolds stress components without affecting \(k\).

\item \textit{The reduction of \(\sigma_n\) is divided into two equal parts, which are respectively added to \(\sigma_t\) and \(\sigma_b\).} We suppose this property by assuming the `two-dimensional isotropy' of the wall-blocking effect on the \(t\)-\(b\) plane. This equal-division assumption can be modified in the future to consider more sophisticated physics.

\item \textit{The shear stress component \(\tau_b\) is preserved.} The nonzeroness of \(\tau_b\) does not directly depend on the existence of walls, so it is natural to assume \(\tau_b\) to be preserved by a wall-blocking correction. It is also worth noting that as long as the realizability constraint (\ref{eq:Rlz}c) is met before the correction, it will always hold if we preserve \(\tau_b\) since both \(\sigma_t\) and \(\sigma_b\) will be increased.

\end{enumerate}
With these properties, the correction methodology can be graphically discussed as follows.

\begin{figure}[htbp]
    \centering
    \includegraphics[width=0.9\linewidth]{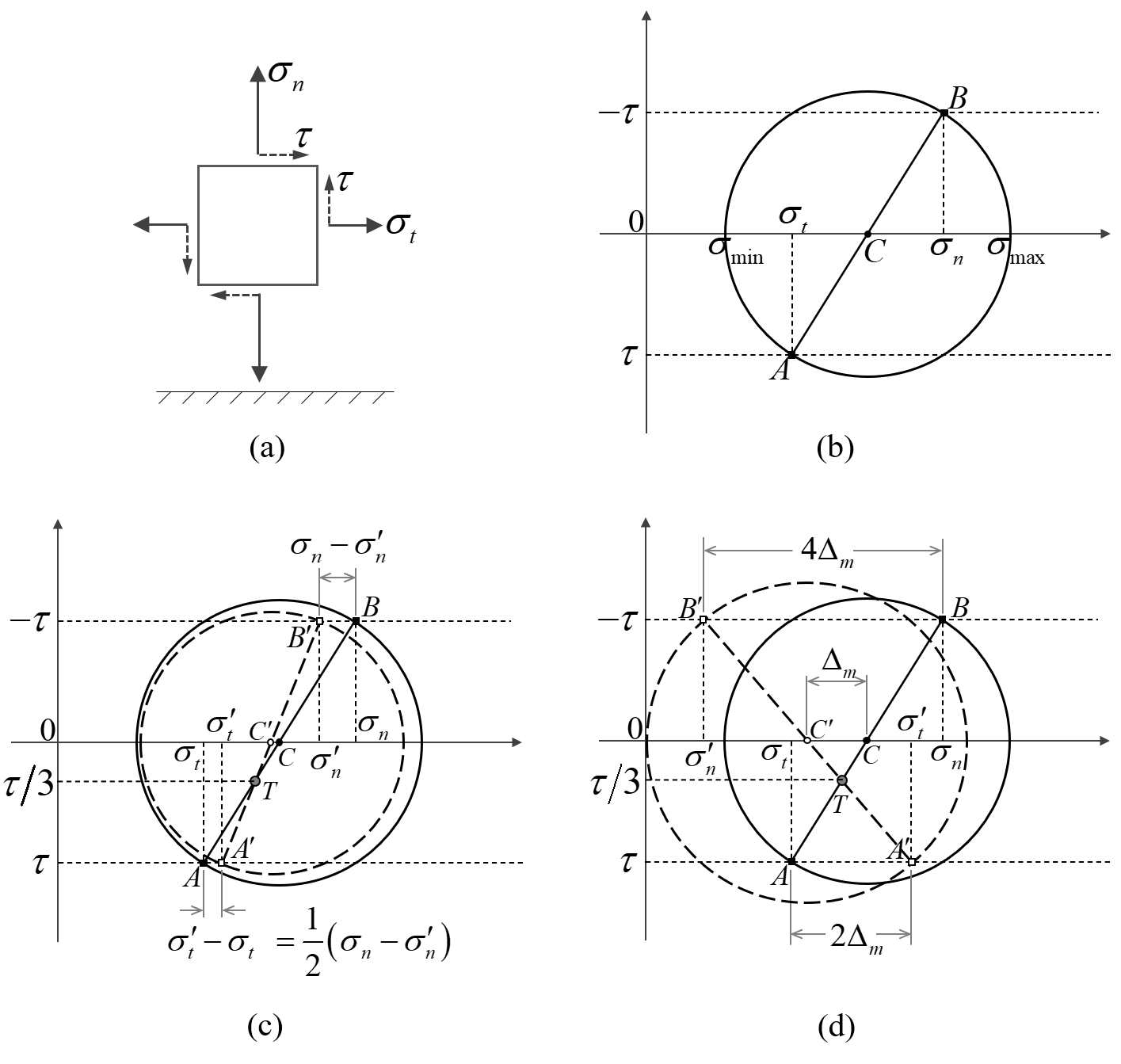}
    \caption{Principle of the Shear-Preserving Wall-Blocking (SPWB) method: (a) Reynolds stress state at a near-wall point; (b) Mohr's circle representation of the stress state; note that \(\tau\leq0\) due to the definition (\ref{eq:tauDef}); (c) alter the EV-predicted stress (represented by the solid Mohr's circle centered at \(C\)) to reduce the wall-normal component while preserving the shear component; a new stress (represented by the dashed Mohr's circle centered at \(C^\prime\)) is thus obtained; and (d) this alteration reaches its limit under the realizability constraint when the new Mohr's circle is tangential to the vertical axis.}
    \label{fig:SPWBprinciple}
\end{figure}

As shown in Fig.~\ref{fig:SPWBprinciple}(c), assume that at a near-wall point, an EV model (e.g.~Eqs.~(\ref{eq:EV}) and (\ref{eq:SSTmodel})) predicts a stress matrix \([\,\widetilde{\textsf{\textbf{R}}}\,]\), which is represented by the solid circle with the diameter line \(\overline{AB}\) and the center \(C\). Given the property (II), the correction should horizontally move the endpoints \(A\) and \(B\) to \(A^\prime\) and \(B^\prime\). To meet the properties (III) and (IV), the two displacements should have the relation \(\overline{AA^\prime}=\overline{B^\prime B}/2\), indicating that these movements are actually a counterclockwise rotation (with a length change) of the diameter line about the point \(T((2\sigma_t+\sigma_n)/3,\tau/3)\). The resulting new (dashed) circle with the diameter line \(\overline{A^\prime B^\prime}\) and the center \(C^\prime\) thus represents the corrected stress matrix \([\,\widetilde{\textsf{\textbf{R}}}^\prime\,]\) with the elements
\begin{subequations}
\label{eq:SPWBcorrection}
\begin{align}
    \sigma_t^\prime & = \sigma_t + 2\Delta\,, \\
    \sigma_n^\prime & = \sigma_n - 4\Delta\,, \\
    \tau^\prime & = \tau \,, \\
    \sigma_b^\prime & = \sigma_b + 2\Delta\,, \\
    \tau_b^\prime & = \tau_b\,,
\end{align}
\end{subequations}
where \(\Delta\equiv\overline{C^\prime C}\) is the displacement of the circle centre, measuring the magnitude of the correction.

As \(\Delta\) increases, the above correction will reach its limit under the realizability constraint (\ref{eq:Rlz}a) when the new Mohr's circle becomes tangential to the vertical axis (i.e.~\(\sigma_\mathrm{min}=0\)), which is illustrated in Fig.~\ref{fig:SPWBprinciple}(d). Through geometrical relations, we obtain the maximum \(\Delta\)
\begin{equation}
\label{eq:DeltaMax}
    \Delta_m = \left[\sigma_n-2\sigma_t+\sqrt{(\sigma_n+2\sigma_t)^2-8\tau^2}\right] / 8\,.
\end{equation}
Given Eqs.~(\ref{eq:SPWBcorrection}) and (\ref{eq:DeltaMax}), the remaining question is to find an appropriate way to preset the field of magnitude \(\Delta\in[0,\Delta_m]\).

\subsubsection{Field of the correction magnitude}
\label{Subsubsec:SPWBrelax}

The wall-blocking effect on Reynolds stress naturally depends on the two length scales: the wall distance \(d_w\) and the turbulent eddy scale \(l\). This paper evaluates \(l\) using the same formulation in the `elliptic relaxation models' \cite{Durbin91} (with an adjusted prefactor) and thus define the dimensionless wall distance \(d_w^-\) by
\begin{subequations}
\label{eq:y_minus}
\begin{align}
    & d_w^- \equiv (\kappa/C_\mu^{3/4})\, (d_w/\,l), \; \kappa=0.41\,, \\
    & l = \max\left\{k^{3/2}/\varepsilon,\;80\,\nu^{3/4}/\varepsilon^{1/4}\right\}.
\end{align}
\end{subequations}
This definition leads to \(d_w^-\approx1\) in the log-layer of a flat-plate boundary layer.

We assume the wall-blocking correction magnitude to be 
\begin{equation}
\label{eq:DeltaRelax}
    \Delta=f_{wb}(d_w^-)\,\Delta_m
\end{equation}
where \(f_{wb}(d_w^-)\in[0,1]\) is a blending function supposed to be monotonically decreasing. As \(d_w^-\to+\infty\), we should have \(f_{wb}(d_w^-)\to0\) due to the far field approximation. For the limit as \(d_w^-\to0\), let us consider the limit of the Eq.~(\ref{eq:SPWBcorrection}b) divided by \(k\):
\begin{equation}
\label{eq:SPWBNdimless}
    \sigma_n^\prime/k = \sigma_n/k - f_{wb}(0) \left[\sigma_n/k\!-\!2\sigma_t/k\!+\!\sqrt{(\sigma_n/k\!+\!2\sigma_t/k)^2\!-\!8(\tau/k)^2}\right] / 2\,.
\end{equation}
As \(d_w^-\to0\), we know that \(k\) tends to \(0\) and \(\varepsilon\) is finite; then the limit of the EV assumption (\ref{eq:EV}a) can be written as
\begin{equation}
\label{eq:EVapprox}
    \textsf{\textbf{R}} / k - 2\,\textsf{\textbf{I}}/3 = - 2\nu_t\,\textsf{\textbf{S}}/k \sim - k\,\textsf{\textbf{S}}/\varepsilon \to 0\,,
\end{equation}
indicating an approximate `isotropic' state. Thus the limit of the right-hand-side of Eq.~(\ref{eq:SPWBNdimless}), in which \(\sigma_t\), \(\sigma_n\) and \(\tau\) are calculated via the EV assumption, is \(2(1-f_{wb}(0))/3\). On the left-hand-side of Eq.~(\ref{eq:SPWBNdimless}), a corrected \(\sigma_n^\prime\) should lead to \(\sigma_n^\prime/k\to0\) in theory. Consequently, we should have \(f_{wb}(0)=1\). 

In this paper, we assume an exponentially decaying function
\begin{equation}
\label{eq:fwb}
    f_{wb}(d_w^-) = e^{-\,d_w^-/\alpha}
\end{equation}
where \(\alpha>0\) is a free parameter. \(\alpha\) controls the scope of near-wall region in which the wall-blocking correction takes effect. If we assume this scope to be comparable with the scope of the log-layer in terms of the order of magnitude, we have \(\alpha\sim\mathcal{O}(1)\).

\subsubsection{Summary}
\label{Subsubsec:SPWBsummary}

Presetting a value of \(\alpha\), we can implement the SPWB correction to Reynolds stress, which are introduced in \S\ref{Subsubsec:tbnCoordSys} - \S\ref{Subsubsec:SPWBrelax}, by following the steps below:

\renewcommand{\theenumi}{(\alph{enumi})}
\begin{enumerate}

\item obtain the pre-correction Reynolds stress \(\textsf{\textbf{R}}\) given by the baseline EV model, e.g.~Eqs.~(\ref{eq:EV}) and (\ref{eq:SSTmodel});

\item calculate the vector bases (\(\textit{\textbf{t}},\textit{\textbf{n}},\textit{\textbf{b}})\) of the local \(t\)-\(n\)-\(b\) system and thus the Jacobian tensor \(\textsf{\textbf{Q}}\) by Eq.~(\ref{eq:tensorQ});

\item transform \(\textsf{\textbf{R}}\) into \(\widetilde{\textsf{\textbf{R}}}\) via Eq.~(\ref{eq:transform});

\item calculate the blending function \(f_{wb}\) by Eqs.~(\ref{eq:y_minus}) and (\ref{eq:fwb});

\item calculate the correction magnitude \(\Delta\) by Eqs.~(\ref{eq:DeltaMax}) and (\ref{eq:DeltaRelax});

\item correct the matrix \([\,\widetilde{\textsf{\textbf{R}}}\,]\) to \([\,\widetilde{\textsf{\textbf{R}}}^\prime\,]\) following Eq.~(\ref{eq:SPWBcorrection});

\item obtain the post-correction Reynolds stress \(\textsf{\textbf{R}}^\prime\) via the inverse transform \(\textsf{\textbf{R}}^\prime = \textsf{\textbf{Q}}\cdot\widetilde{\textsf{\textbf{R}}}^\prime\cdot\textsf{\textbf{Q}}^\mathrm{T}\).

\end{enumerate}

\subsection{Double-Scale Double-Linear-eddy-viscosity (DSDL) model}
\label{Subsec:DSDL}

The conceptual DSDL model was proposed by \citet{Hao21} with its original objective of improving the predictions of \(k\) and other turbulent quantities around the large separation regions behind bluff bodies. In this paper, it is directly employed to address the EV models' general underpredictions of scalar transport behind a structure. This subsection briefly introduces the DSDL model's basic idea and formulation, using \(k\)-\(\omega\) SST model (Eqs.~(\ref{eq:EV}) and (\ref{eq:SSTmodel})) as the baseline.

\begin{figure}[htbp]
    \centering
    \includegraphics[width=1.0\linewidth]{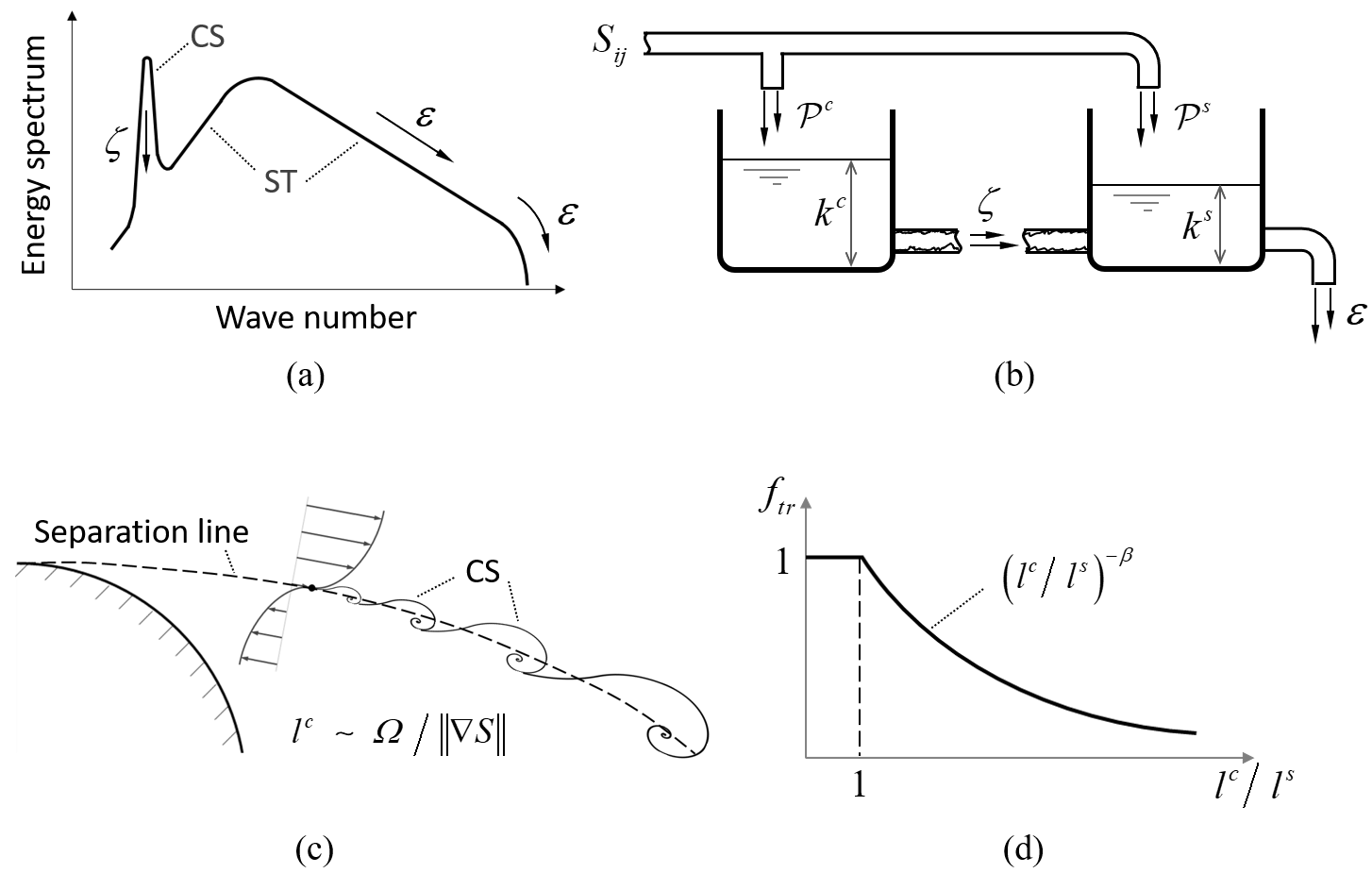}
    \caption{Principle of the Double-Scale Double-Linear-eddy-viscosity (DSDL) model: (a) a sketch of turbulence energy spectrum distorted by coherent structure; (b) an analogy between the spectral energy transfer and the flow in a tank-tube system; (c) the supposition of the CS length scale, which naturally yields intense generation of \(k^c\) around the separation line behind a structure; and (d) the assumed power function to indicate the effect of length scale ratio on the CS-ST energy transfer.}
    \label{fig:DSDLprinciple}
\end{figure}

First, the DSDL model intends to distinguish the fluctuations induced by vortex shedding behind bluff bodies, a typical form of coherent structures (CS, or superscript \(c\)), from the background `stochastic' turbulence (ST, or superscript \(s\)). Both CS and ST contribute to the total Reynolds stress \(\bm{\mathsf{R}}=\bm{\mathsf{R}}^c+\bm{\mathsf{R}}^s\) and the total fluctuation kinetic energy \(k=k^c+k^s\). Fig.~\ref{fig:DSDLprinciple}(a) is a sketch of the energy spectrum of turbulence embedding CS, where \(\zeta\) represents the energy transfer rate (ETR) from \(k^c\) to \(k^s\). The variables \(k^c\), \(k^s\) and \(\omega\equiv \varepsilon/(C_\mu k^s)\) are assumed to be governed by
\begin{subequations}
\label{eq:energies}
\begin{align}
    & \bar{D}_t k^c = \mathcal{P}^c - \zeta + \nabla\cdot\left[\, (\nu+\sigma_k\nu_t^c) \nabla k^c \,\right], \\
    & \bar{D}_t k^s = \mathcal{P}^s_\mathrm{lim} + \zeta - \varepsilon + \nabla\cdot\left[\, (\nu+\sigma_k\nu_t^s) \nabla k^s \,\right], \\
    & \bar{D}_t\omega = \left[\,\alpha(\mathcal{P}^s_\mathrm{lim}+\zeta) - (\beta/C_\mu)\,\varepsilon\,\right]/\nu_t^s + \nabla\cdot\left[\,(\nu+\sigma_\omega\nu_t^s)\nabla\omega\,\right] + C\!D\,, \\
    & \mathcal{P}^s_\mathrm{lim} = \mathrm{min}\{\mathcal{P}^s,\, 10\,\varepsilon\},
\end{align}
\end{subequations}
where we have two eddy viscosities \(\nu_t^c\) and \(\nu_t^s\) and two energy production rates \(\mathcal{P}^c\equiv-\textsf{\textbf{R}}^c:\textsf{\textbf{S}}\) and \(\mathcal{P}^s\equiv-\textsf{\textbf{R}}^s:\textsf{\textbf{S}}\). (Note that in this paper we drop the Kato's production modification that was used in \cite{Hao21} to avoid the stagnation point anomaly.) The energy budget indicated by Eq.~(\ref{eq:energies}) can be analogized to a tank-tube system in Fig.~\ref{fig:DSDLprinciple}(b).

Second, the DSDL model assumes \(\textsf{\textbf{R}}^c\) and \(\textsf{\textbf{R}}^s\) to respectively follow the EV assumption. For ST, \(\textsf{\textbf{R}}^s\) can be determined similarly to Eq.~(\ref{eq:EV}):
\begin{subequations}
\label{eq:EVsto}
\begin{align}
    & \textsf{\textbf{R}}^s = 2k^s\,\textsf{\textbf{I}}/3 - 2\nu_t^s\,\textsf{\textbf{S}}\,, \\
    & \nu_t^s = k^s\,/\,\mathrm{max}(\omega,SF_2/a_1)\, .
\end{align}
\end{subequations}
For CS, the model supposes an algebraic form for the length scale \(l^c\):
\begin{equation}
\label{eq:lengthCS}
    l^c = (\kappa/C_\mu^{3/4})\,\min\left\{\mathit{\Omega}/\|\nabla S\|, \; d_w\left(1-e^{-d_w^+/A^+}\right)\right\},\quad \mathit{\Omega}\equiv\sqrt{2\,\bm{\mathsf{\Omega}}:\bm{\mathsf{\Omega}}}
\end{equation}
where \(\bm{\mathsf{\Omega}}\equiv (\nabla\textit{\textbf{U}}-\left(\nabla\textit{\textbf{U}}\right)^\mathrm{T})/2\), and \(d_w^+=C_\mu^{1/4}(k^s)^{1/2}d_w/\nu\); then \(\textsf{\textbf{R}}^c\) is determined by
\begin{subequations}
\label{eq:EVcoh}
\begin{align}
    & \textsf{\textbf{R}}^c = 2k^c\,\textsf{\textbf{I}}/3 - 2\nu_t^c\,\textsf{\textbf{S}}\,, \\
    & \nu_t^c = C_\mu(k^c)^{1/2}l^c\, .
\end{align}
\end{subequations}
The length scale \(l^c\) evaluated by (\ref{eq:lengthCS}) can be large near the center of a strong shear layer where \(\mathit{\Omega}\) is large but \(\parallel\nabla S\parallel\) is small. This property yields intense generation of \(k^c\) in these regions, which are generally induced by the large flow separations behind structures, as shown in Fig.~\ref{fig:DSDLprinciple}(c).

Lastly, based on a dimensional analysis and an analogy with the `return-to-isotropy' evolution of Reynolds stress anisotropy, the ETR \(\zeta\) is parameterized as
\begin{equation}
\label{eq:ETRform}
    \zeta = c_{tr}\; f_{tr}(l^c/l^s)\;(k^c/k)\;.
\end{equation}
where \(c_{tr}\in(1,\infty)\) is a free parameter. \(c_{tr}\) controls the intrinsic rate of CS-ST energy transfer, and can be analogized to the `cross-section area' of the tube connecting the two tanks in Fig.~\ref{fig:DSDLprinciple}(b). In Eq.~(\ref{eq:ETRform}), the ST length scale \(l^s\) is determined similarly to Eq.~(\ref{eq:y_minus}b):
\begin{equation}
\label{eq:lengthST}
    l^s = \max\left\{(k^s)^{3/2}/\varepsilon,\;80\,\nu^{3/4}/\varepsilon^{1/4}\right\},
\end{equation}
and \(f_{tr}(l^c/l^s)\in[0,1]\) is a monotonically non-increasing function representing the effect of length scale ratio \(l^c/l^s\) on ETR. In this paper, we adopt the following power function as shown in Fig.~\ref{fig:DSDLprinciple}(d) (which is different from the smoothed step function in \cite{Hao21}):
\begin{equation}
\label{eq:Ftrans}
    f_{tr}(l^c/l^s) = \max\{l^c/l^s,\,1\}^{-\beta}
\end{equation}
where \(\beta\in[0,\infty)\) is a second free parameter besides \(c_{tr}\) of Eq.~(\ref{eq:ETRform}). \(\beta\) controls the CS-ST scale separation effect on ETR, and can be analogized to the `inner-surface roughness' of the tube in Fig.~\ref{fig:DSDLprinciple}(b). When \(c_{tr}\to\infty\) and \(\beta\) is finite, the DSDL model is reduced to the baseline single-scale EV model.

When the DSDL model is applied jointly with the scalar flux model (\ref{eq:GGDHmodel}), \(\nu_t\) in (\ref{eq:GGDHmodel}) is supposed to be \(\nu_t^c+\nu_t^s\).

\subsection{The Reynolds-Stress-Shape-Perturbation (RSSP) for comparison}
\label{Subsec:RSSP}

In addition to SPWB and/or DSDL, for the sake of comparison, we separately apply a method that was previously proposed by \cite{Emory13,Gorle12} for correcting EV models. We term it the `Reynolds-Stress-Shape-Perturbation (RSSP)' method. It has exhibited promising performance in terms of model bias reductions and bounding behaviors as an uncertainty quantification (UQ) approach in a broad range of problems (e.g.~\cite{Iaccarino17,Gorle19,Hao20a,Xiao16}). This subsection briefly reviews the RSSP method.

A Reynolds stress \(\bm{\mathsf{R}}\) can be eigen-decomposed as
\begin{equation}
\label{eq:eigDecp}
    \bm{\mathsf{R}} = 2k\left(\,\bm{\mathsf{I}}/3+\bm{\mathsf{V}}\cdot \bm{\mathsf{\Lambda}}\cdot\bm{\mathsf{V}}^\mathrm{T}\,\right)
\end{equation}
where the tensor \(\bm{\mathsf{\Lambda}}\) has a diagonal matrix \([\,\bm{\mathsf{\Lambda}}\,]=\mathrm{diag}[\,\lambda_1,\lambda_2,\lambda_3\,]\), with \(\lambda_1+\lambda_2+\lambda_3=0\) and \(\lambda_1\geq\lambda_2\geq\lambda_3\), representing the shape of \(\bm{\mathsf{R}}\); and \(\bm{\mathsf{V}}\) is an orthogonal tensor representing the orientation of \(\bm{\mathsf{R}}\). Under the realizability constraint, we have three limit shapes: the one-component limit with \([\,\bm{\mathsf{\Lambda}}_\mathrm{C1}\,]=\mathrm{diag}[\,2/3,-1/3,-1/3\,]\), the two-component limit with \([\,\bm{\mathsf{\Lambda}}_\mathrm{C2}\,]=\mathrm{diag}[\,1/6,1/6,-1/3\,]\), and the three-component (i.e.~isotropic) limit with  \([\,\bm{\mathsf{\Lambda}}_\mathrm{C3}\,]=\mathrm{diag}[\,0,0,0\,]\).

The RSSP method states that to estimate the bounds of a QoI, one can simply preserve the orientation \(\bm{\mathsf{V}}\) of the Reynolds stress \(\bm{\mathsf{R}}\) resulting from a baseline EV model (e.g.~Eqs.~(\ref{eq:EV}) and (\ref{eq:SSTmodel})) while perturb the shape \(\bm{\mathsf{\Lambda}}\) linearly towards \(\bm{\mathsf{\Lambda}}_\mathrm{C1}\), \(\bm{\mathsf{\Lambda}}_\mathrm{C2}\) and \(\bm{\mathsf{\Lambda}}_\mathrm{C3}\) respectively. The corrected Reynolds stress \(\bm{\mathsf{R}}^\prime\) is thus expressed as
\begin{subequations}
\label{eq:RSSPcorrection}
\begin{align}
    & \bm{\mathsf{R}}^\prime = 2k\left(\,\bm{\mathsf{I}}/3+\bm{\mathsf{V}}^\prime\cdot \bm{\mathsf{\Lambda}}^\prime\cdot\bm{\mathsf{V}}^{\prime\,\mathrm{T}}\,\right), \\
    & \bm{\mathsf{\Lambda}}^\prime = (1-m)\,\bm{\mathsf{\Lambda}} + m\,\bm{\mathsf{\Lambda}}_\mathrm{C1/C2/C3}\,, \quad \bm{\mathsf{V}}^\prime = \bm{\mathsf{V}}\,,
\end{align}
\end{subequations}
where \(m\in[0,1]\) is a free parameter controlling the perturbation magnitude.

\cite{Emory13} and \cite{Gorle12} proved that among all the routes of perturbing \(\bm{\mathsf{\Lambda}}\) under the realizability constraint, the perturbation linearly towards \(\bm{\mathsf{\Lambda}}_\mathrm{C1}\) (or \(\bm{\mathsf{\Lambda}}_\mathrm{C3}\)) as Eq.~(\ref{eq:RSSPcorrection}b) is the most effective route to increase (or decrease) the shear stress magnitude in a parallel shear turbulence as well as the energy production rate \(\mathcal{P}\); these two factors directly influence the general level of mean momentum transport. Therefore, it is reasonable to some extent to expect the results of the perturbations towards \(\bm{\mathsf{\Lambda}}_\mathrm{C1}\) and \(\bm{\mathsf{\Lambda}}_\mathrm{C3}\) to appropriately bound a QoI related to momentum transport.


In this paper, we determine the perturbation magnitudes \(m\) in the same way as provided in \cite{Hao20a}: for each simulation RSSP-C\(n\) (\(n\in\{1,2,3\}\)), a uniform \(m\) is set for all Reynolds stresses in the entire domain; the value of \(m\) is selected to be \(\min\{0.5,m_c\}\) where \(m_c\) is the maximum \(m\) that can result in a well converged steady-state solution.

\section{Results and discussions}
\label{Sec:Applications}

In this section, we apply the EV model correction methodologies introduced in \S\ref{Sec:Methods} to two test cases. The simulations are divided into five categories as listed in table \ref{tab:MethodSummary}. Only steady-state solutions such that \(\partial_t=0\) for all the solved equations in table \ref{tab:MethodSummary} are concerned.

\begin{table}
\renewcommand\arraystretch{1.5}
  \footnotesize
  \begin{center}
\def~{\hphantom{0}}
    \begin{tabular}{p{0.16\linewidth}p{0.38\linewidth}p{0.37\linewidth}}
        \hline
        Denotation & Free Parameter(s) & Equations Solved \\
        \hline
        Baseline & N/A & (\ref{eq:basicRANS})-(\ref{eq:GGDHmodel}) \\
        RSSP-C\(n\) & \(n\in\{1,2,3\}\), \(m\in[0,1]\) & (\ref{eq:basicRANS})-(\ref{eq:GGDHmodel}),(\ref{eq:eigDecp})-(\ref{eq:RSSPcorrection}) \\
        SPWB & \(\alpha\sim\mathcal{O}(1)\) & (\ref{eq:basicRANS})-(\ref{eq:GGDHmodel}) \& steps in \S\ref{Subsubsec:SPWBsummary} \\
        DSDL & \(c_{tr}\in(1,\infty)\), \(\beta\in[0,\infty)\) & (\ref{eq:basicRANS}),(\ref{eq:GGDHmodel}),(\ref{eq:energies})-(\ref{eq:Ftrans}) \\
        SPWB-DSDL & \(\alpha\sim\mathcal{O}(1)\), \(c_{tr}\in(1,\infty)\), \(\beta\in[0,\infty)\) & (\ref{eq:basicRANS}),(\ref{eq:GGDHmodel}),(\ref{eq:energies})-(\ref{eq:Ftrans}) \& steps in \S\ref{Subsubsec:SPWBsummary} \\
        \hline
    \end{tabular}
    \caption{Categories of steady-RANS simulations implemented in \S\ref{Sec:Applications}}
    \label{tab:MethodSummary}
  \end{center}
\end{table}

All steady-RANS simulations are implemented using the open source software OpenFOAM \cite{Weller98}. The differential equations are spatially discretized using the second-order finite volume method. The SIMPLE algorithm is adopted for pressure-velocity decoupling. The solution is considered to be converged when the global residuals of all transport equations decrease more than four orders of magnitude and the QoIs reach their steady-state values. The near-wall grid resolutions are high enough to ensure \(\mathrm{\Delta} y^+\leq1\), avoiding the use of wall functions. For each case, the results presented in this section were demonstrated to be grid independent by performing simulations with a finer mesh (resolution \(25\%\sim30\%\) higher in each dimension) and verifying that discrepancies between the QoIs predicted by both meshes were smaller than 1\%.

\subsection{Forced heat convection in a pin-fin array}
\label{Subsec:PinFinArray}

\subsubsection{Flow configuration, reference database, and simulation set-ups}
\label{Subsubsec:PinFinConfiguration}

The first test case is forced heat convection in a pin-fin array. This case was first studied experimentally by \cite{Ames05}, followed by several numerical studies including a high-fidelity large eddy simulation (LES) reported by \cite{Hao19}. We will use the validated LES results presented in \cite{Hao19} as the reference data in this subsection.

\begin{figure}[htbp]
    \centering
    \includegraphics[width=0.85\linewidth]{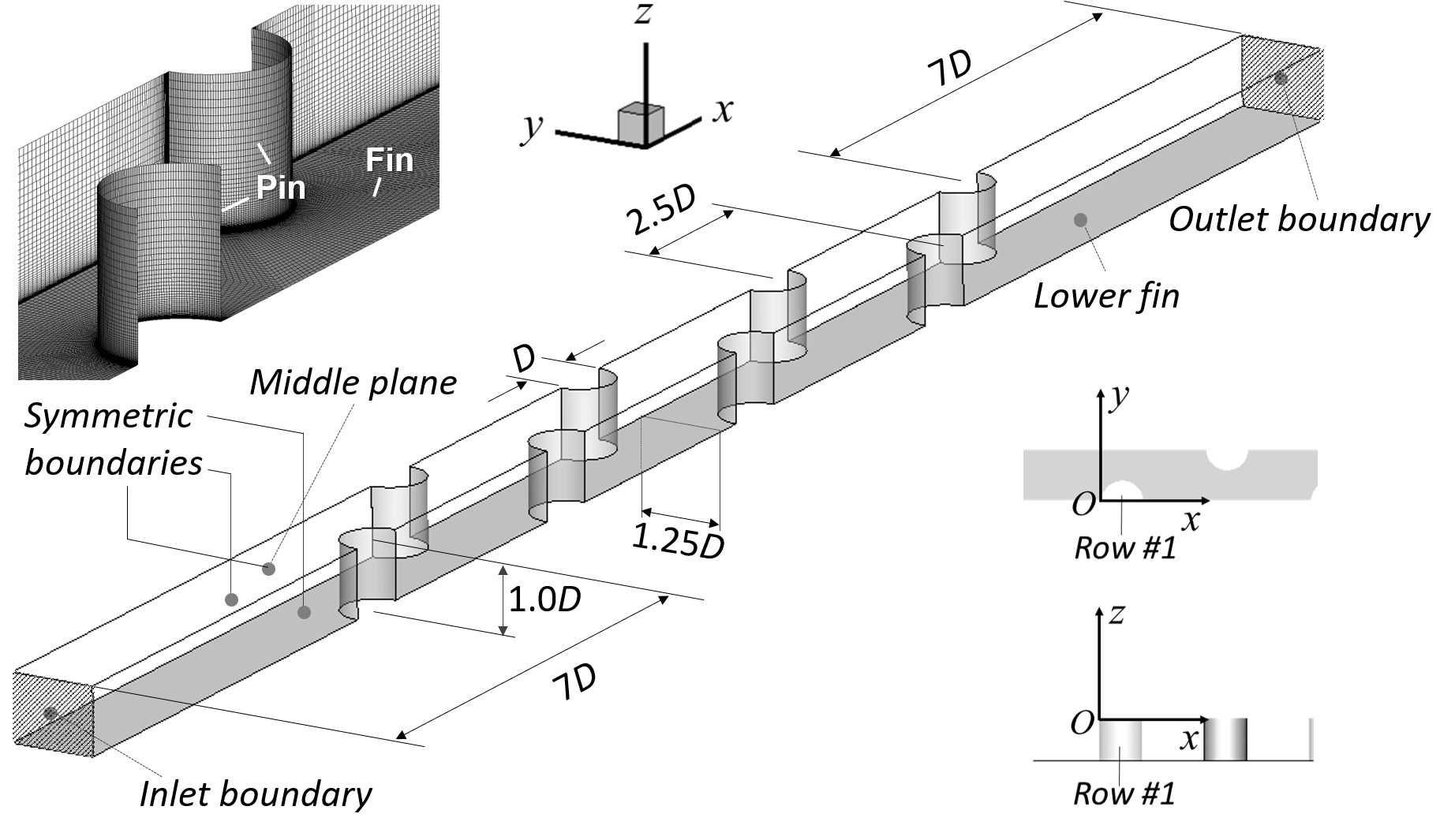}
    \caption{Computational domain and mesh of the pin fin array case}
    \label{fig:PinFin_Configuration}
\end{figure}

Fig.~\ref{fig:PinFin_Configuration} shows the computational domain for this pin-fin array case, ranging over a half of the lateral (\(y\)) pin spacing and a half of the fin spacing due to symmetries. The configuration consists of eight staggered rows of circular cylinders (pins) with diameter $D$ placed between two parallel flat plates (fins). Both streamwise (\(x\)) and lateral spacings of the pins are $2.5D$, and distance between the fins is $2D$. At inlet, fluids with a high temperature $\mathit{\Theta}_{in}$ and low turbulence intensity $\mathrm{Tu}_{in}=1.5\%$ flow into the array. Turbulent forced heat convection occurs between the fluids and all the surfaces of pins and fins with a constant temperature $\mathit{\Theta}_w$ that is lower than $\mathit{\Theta}_{in}$. The Reynolds number is $\mathrm{Re}=V_m D/\nu=10,000$ where $V_m$ is the average velocity on the throat cross section between lateral adjacent pins, and the Prandtl number is $\mathrm{Pr=\nu/\kappa}=0.71$. The entire computational domain is discretized using 2.65 million hexahedral cells. More details on the simulation set-ups can be found in \cite{Hao20b}.

In this subsection, the parameter \(m\) values for RSSP-C1, RSSP-C2 and RSSP-C3 are 0.5, 0.5 and 0.2, respectively. The parameter \(c_{tr}\) for DSDL and SPWB-DSDL is fixed to be $1.6$, a value in the range \(1.5\sim1.8\) that was suggested by \cite{Hao21} for bluff body flows; effect of various \(\beta\) values on results will be examined.

In this case, the flow past one of the pins near the middle plane $z=0$ have two featured flow patterns: an upstream impingement pattern and a downstream quasi-two-dimensional separation pattern. In the remainder of this subsection, we will primarily investigate the flow and heat transfer related to these two features; in particular, since the features change little from the \(4^\mathrm{th}\) to the \(7^\mathrm{th}\) row of pins where turbulence is fully developed (see \cite{Hao19,Hao20a,Hao20b}), we focus our discussion on the results around the \(5^\mathrm{th}\) row of pins. The heat transfer on the fins will be discussed secondarily.

\subsubsection{Flow-related results}
\label{Subsubsec:PinFinResultsFlow}

\begin{figure}[htbp]
    \centering
    \includegraphics[width=1.0\linewidth]{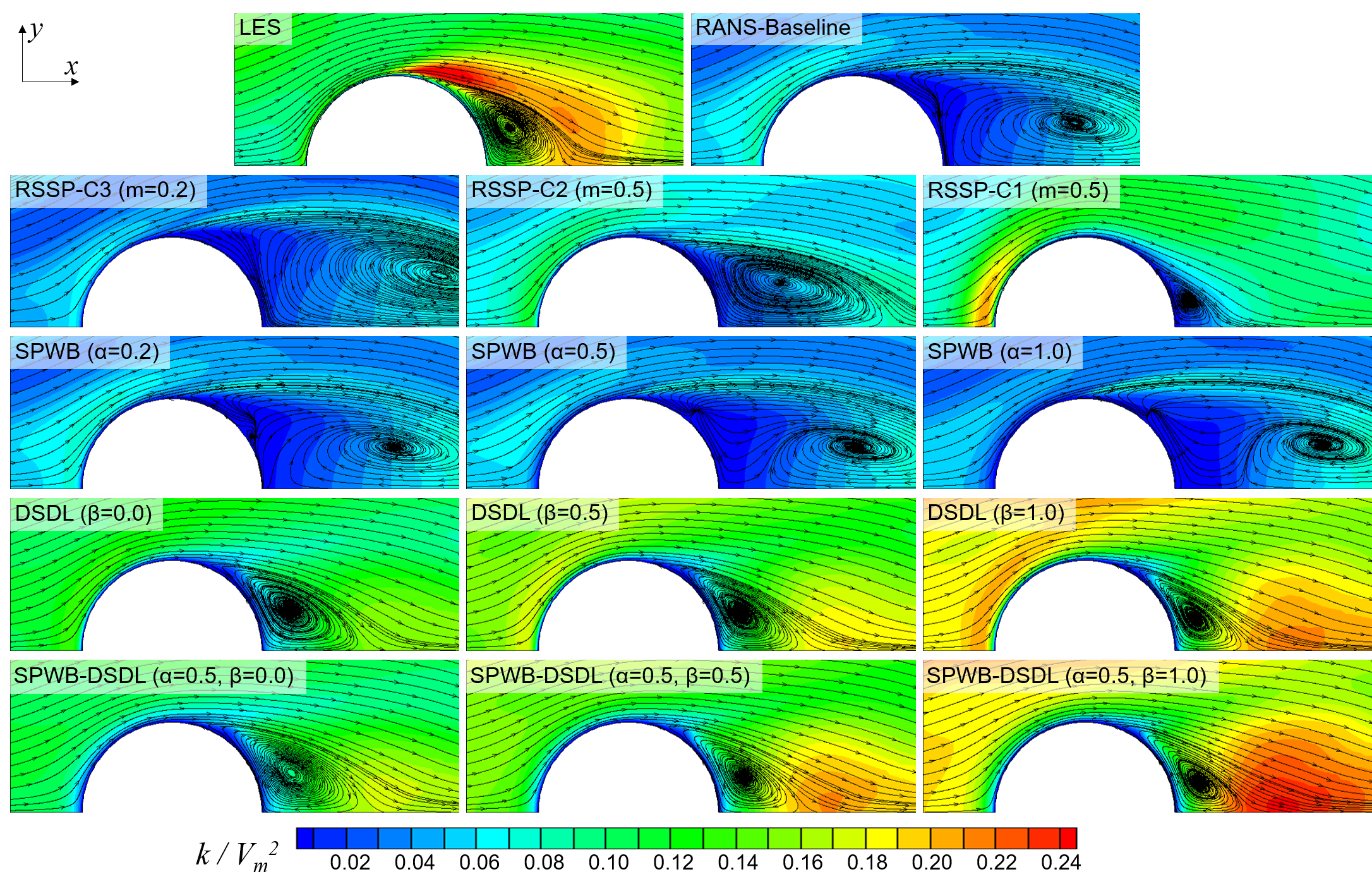}
    \caption{Mean streamlines and contours of turbulence kinetic energy \(k\) on the middle plane \(z=0\) around the \(5^\mathrm{th}\) row of pins}
    \label{fig:PinFin_Row5_ktot_stream}
\end{figure}

Fig.~\ref{fig:PinFin_Row5_ktot_stream} shows mean streamlines and \(k\) contours on the middle plane \(z=0\) around the \(5^\mathrm{th}\) row of pins. Compared to the LES result, the baseline model considerably overestimates the size of recirculation region behind the pin, and significantly underestimates the energy level in and near that region. The RSSP-C2 and RSSP-C1 corrections to the baseline model effectively decrease the recirculation size, but little improve the energy prediction; for the upwind side they even aggravate the stagnation point anomaly. The SPWB corrections to the baseline model lower the energy level near stagnation point while little affect either size or energy of recirculation region. The DSDL model with various \(\beta\) values predicts a set of recirculation sizes that are well consistent with the LES result; increasing \(\beta\) tends to increase the energy level in and downstream of recirculation region as well as near stagnation point. The combination of DSDL with the SPWB alleviates the former's stagnation point anomaly, and meanwhile slightly raises the energy level downstream of the pin. To summarize, the combined model SPWB-DSDL substantially improves upon the baseline EV model in quality in terms of predicting the size of and the energy level near recirculation region, while avoiding the stagnation point anomaly. One remaining problem is on the region near separation onset point, where the energy is still apparently underestimated.

\begin{figure}[htbp]
    \centering
    \includegraphics[width=1.0\linewidth]{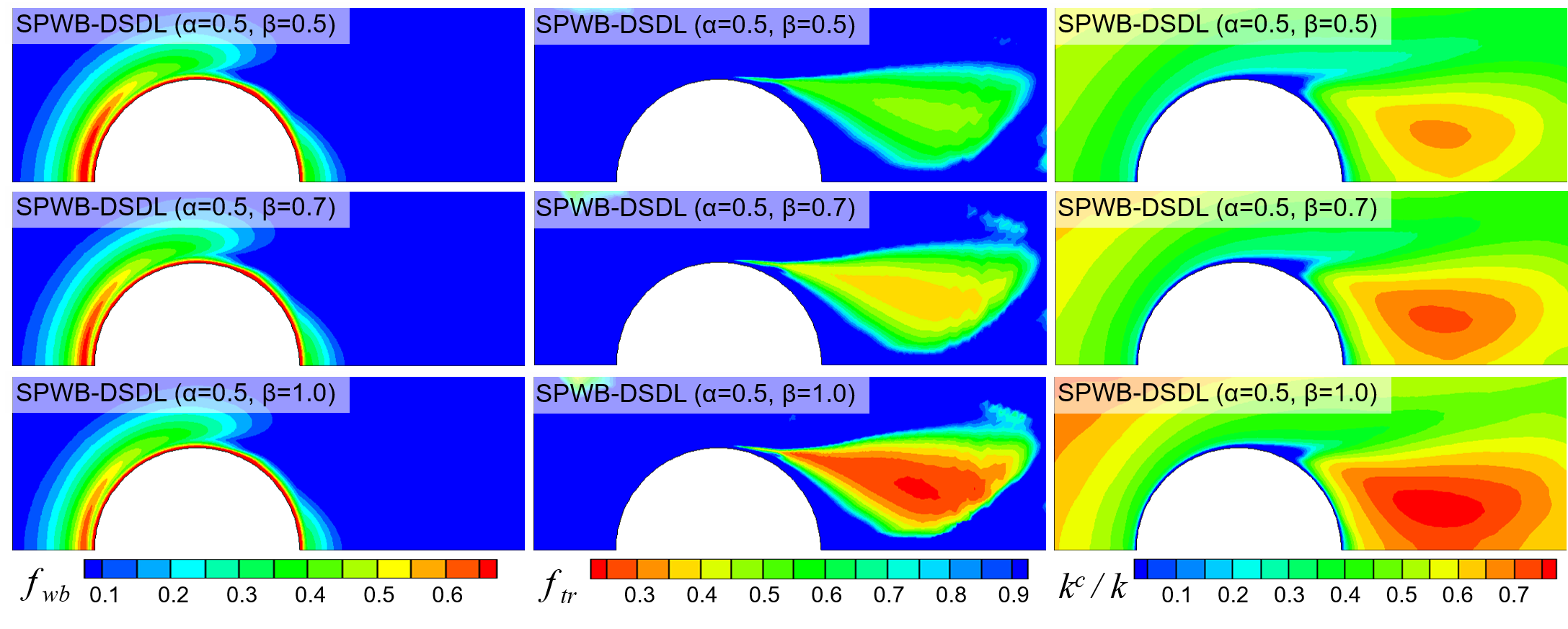}
    \caption{Contours of wall blocking function \(f_{wb}\) (left column), energy transfer function \(f_{tr}\) (middle column), and CS energy fraction \(k^c/k\) (right column) for three SPWB-DSDL simulations respectively with \(\beta=0.5\) (upper row), \(0.7\) (middle row) and \(1.0\) (bottom row)}
    \label{fig:PinFin_Row5_Blending2}
\end{figure}

Fig.~\ref{fig:PinFin_Row5_Blending2} illustrates the mechanisms of SPWB-DSDL model by showing the contours of wall-blocking function \(f_{wb}\) (in Eq.~(\ref{eq:DeltaRelax})), energy transfer function \(f_{tr}\) (defined by Eq.~(\ref{eq:ETRform})), and CS energy fraction \(k^c/k\). The \(f_{wb}\) contours indicate that the wall-blocking correction takes effect apparently more in the regions near stagnation point and upwind surface than in the region near downwind surface. In each \(f_{tr}\) contour, a fan-shaped region can be identified, with its apex at the separation onset point and its width gradually dilating along the separation line. This region has small \(f_{tr}\) and thus large \(l^c/l^s\), indicating that \(k^c\) is intensely generated and the energy transfer from \(k^c\) to \(k^s\) is suppressed there. Interestingly, the SPWB-DSDL model does predict that the separation onset immediately induces a local low \(f_{tr}\) region, which continuously dilates downstream; however, the model fails to predict a high level of \(k^c/k\) or \(k\) (Fig.~\ref{fig:PinFin_Row5_ktot_stream}) immediately near the separation onset point due to the insufficient historical accumulation of \(k^c\) there. This failure implies that an improved model in the future could consider a much stronger \(k^c\) production near the separation onset point without increasing the shear stress that controls the separation onset itself.

\begin{figure}[htbp]
    \centering
    \includegraphics[width=1.0\linewidth]{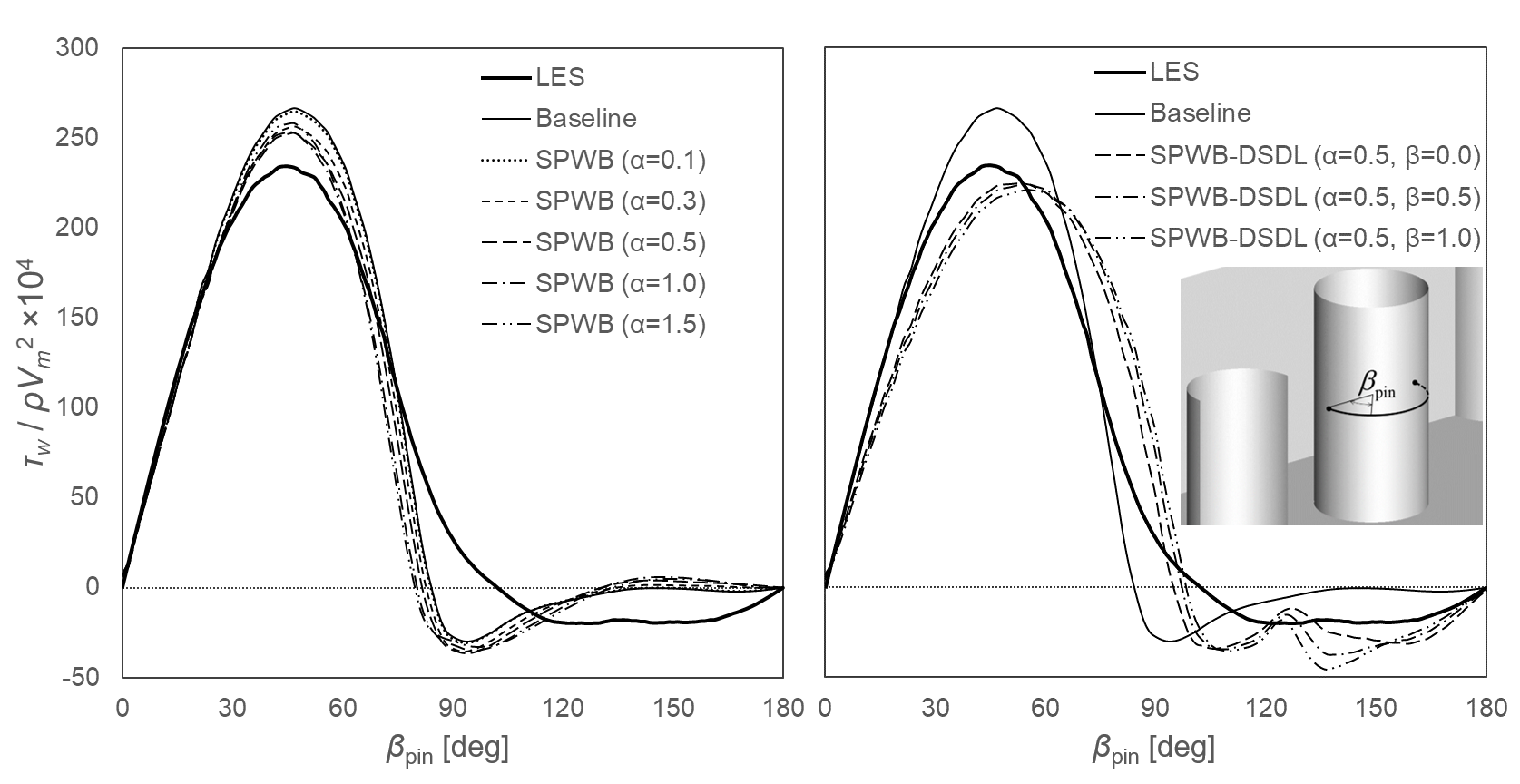}
    \caption{Friction coefficient distributions on surface of the \(5^\mathrm{th}\) row of pins}
    \label{fig:PinFin_Row5_Friction}
\end{figure}

Fig.~\ref{fig:PinFin_Row5_Friction} shows the friction stress distributions on the pin surface predicted by SPWB and SPWB-DSDL. The baseline model predicts a maximum friction value and a separation onset location that are respectively 14\% higher than and \(18^\circ\) ahead of the LES results. The SPWB corrections, with a variety of \(\alpha\) up to 1.5, have fairly limited effect on the surface friction thanks to the shear-preserving design. They change the baseline model results of maximum friction value and separation onset location by less than 5\% and by smaller than \(5^\circ\), respectively. The SPWB-DSDL model, with \(\beta\) from 0 to 1, moderately improves the predictions of the two properties. The errors relative to LES results are \(4\%\sim5\%\) and \(4^\circ\sim7^\circ\).

\begin{figure}[htbp]
    \centering
    \includegraphics[width=1.0\linewidth]{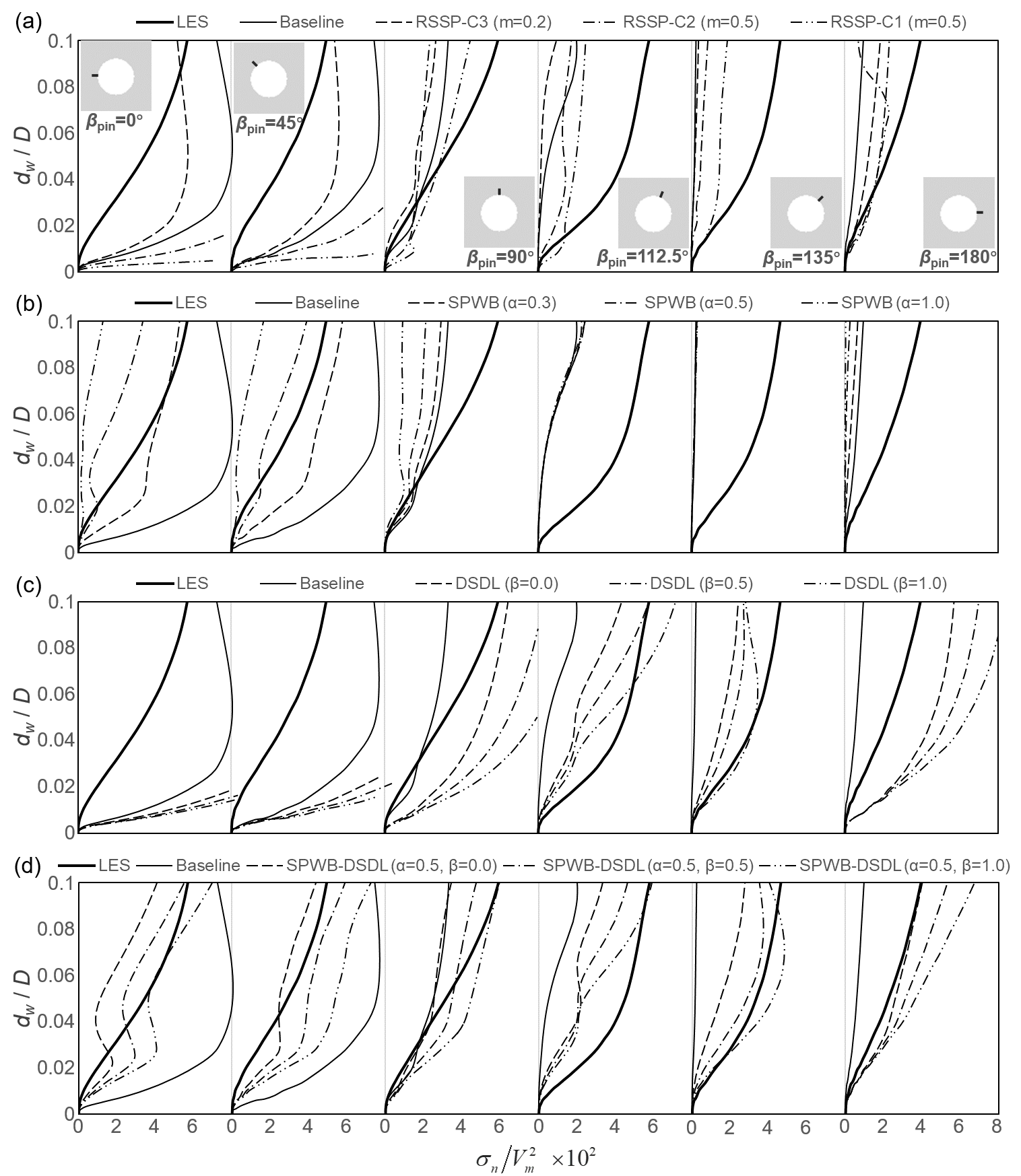}
    \caption{Near-wall profiles of wall-normal Reynolds stress \(\sigma_n\) versus wall distance \(d_w\) at six different locations (\(\beta_\mathrm{pin}=0^\circ, 45^\circ, 90^\circ, 112.5^\circ, 135^\circ \textup{ and } 180 ^\circ\) with \(z=0\) for all) on surface of the \(5^\mathrm{th}\) row of pins}
    \label{fig:PinFin_Row5_WNstress}
\end{figure}

Fig.~\ref{fig:PinFin_Row5_WNstress} plots the near-wall profiles of wall-normal stress \(\sigma_n\), which has strong impact on heat transfer prediction as discussed in \S\ref{Subsubsec:SPWBmotive}, at six different locations on the pin surface. At \(\beta_\mathrm{pin}=0^\circ\) and \(45^\circ\) where the boundary layer is subjected to strong mean flow contraction in wall-normal direction, all the simulations without SPWB correction, i.e.~Figs.~\ref{fig:PinFin_Row5_WNstress}(a,c), significantly overpredict the level of near-wall \(\sigma_n\). At \(\beta_\mathrm{pin}=135^\circ\) and \(180^\circ\) where the wall is covered by recirculation flow, the simulations without DSDL model, i.e.~Figs.~\ref{fig:PinFin_Row5_WNstress}(a,b), mostly underpredict \(\sigma_n\). In general for the whole range of \(\beta_\mathrm{pin}\), the SPWB-DSDL simulations with \(\alpha=0.5\) and \(\beta=0\sim1\), i.e.~Fig.~\ref{fig:PinFin_Row5_WNstress}(d), predict more reasonable near-wall \(\sigma_n\) levels than the other categories of simulations.

\subsubsection{Scalar-related results}
\label{Subsubsec:PinFinResultsScalar}

\begin{figure}[htbp]
    \centering
    \includegraphics[width=1.0\linewidth]{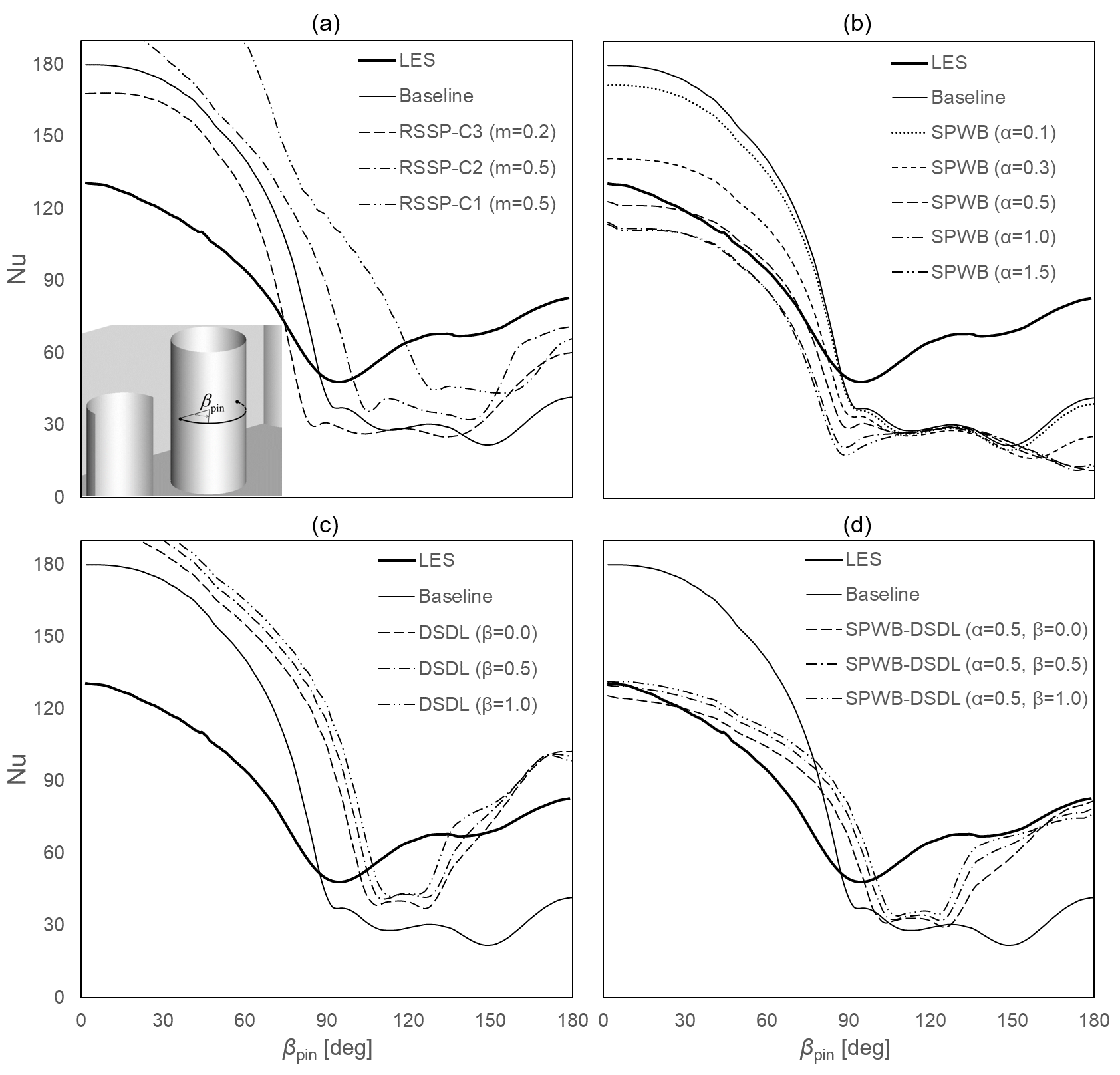}
    \caption{Nusselt number distributions on surfaces of the \(5^\mathrm{th}\) row of pins}
    \label{fig:PinFin_Row5_NusseltPin}
\end{figure}

Define the local Nusselt number at a pin surface point as
\begin{equation}
\label{eq:NusseltDef}
    \mathrm{Nu}=(\partial_n\mathit{\Theta})_w \, D \,/\, (\mathit{\Theta}_b-\mathit{\Theta}_w)
\end{equation}
where \((\partial_n\mathit{\Theta})_w\) is wall-normal gradient of \(\mathit{\Theta}\) on the wall and \(\mathit{\Theta}_b\) is the bulk temperature calculated by flow-rate-weighted averaging \(\mathit{\Theta}\) in the space with \(4.25D\leq x \leq6.75D\). Fig.~\ref{fig:PinFin_Row5_NusseltPin} shows the \(\mathrm{Nu}\) distributions on the pin surface.

In Fig.~\ref{fig:PinFin_Row5_NusseltPin}(a), compared to the LES result, the baseline model considerably overpredicts \(\mathrm{Nu}\) on the upwind surface. At stagnation point the overprediction is 38\%. This number is even increased to 115\% and 57\% by RSSP-C1 and RSSP-C2, while only slightly decreased to 29\% by RSSP-C3. By contrast, on the downwind surface the baseline model significantly underpredicts \(\mathrm{Nu}\), with an error of 50\% at \(\beta_\mathrm{pin}=180^\circ\). This number is moderately decreased to \(14\%\sim27\%\) by the three RSSP simulations. In general, none of the three RSSP simulations effectively correct the baseline EV model's upwind overprediction or downwind underprediction of \(\mathrm{Nu}\) on the pin surface.

Fig.~\ref{fig:PinFin_Row5_NusseltPin}(b) shows the \(\mathrm{Nu}\) distributions predicted by SPWB corrections with various \(\alpha\). The increase of \(\alpha\) from 0.1 to 1.5 apparently decreases \(\mathrm{Nu}\) on the upwind surface and on a small part of the downwind surface with \(\beta_\mathrm{pin}=155^\circ\sim180^\circ\), while little changes \(\mathrm{Nu}\) on the surface with \(\beta_\mathrm{pin}=105^\circ\sim150^\circ\) that is just downstream of the separation point. The results of \(\mathrm{Nu}\) on the entire surface tend to converge as \(\alpha>1.0\), leading to a limit stagnation point \(\mathrm{Nu}\) that is 13\% lower than the LES result.

Fig.~\ref{fig:PinFin_Row5_NusseltPin}(c) shows the \(\mathrm{Nu}\) distributions predicted by DSDL model with various \(\beta\). The DSDL model predicts higher \(\mathrm{Nu}\) than baseline model on the entire pin surface, especially for the recirculation region with \(\beta_\mathrm{pin}=130^\circ\sim180^\circ\). At \(\beta_\mathrm{pin}=180^\circ\), the DSDL predicted values of \(\mathrm{Nu}\) are \(19\%\sim24\%\) higher than the LES result. As \(\beta\) changes from 0 to 1, the absolute changes of DSDL predicted \(\mathrm{Nu}\) are around 10, a relatively small amount compared to the absolute discrepancies between the DSDL and the baseline model results, which are around 40.

The results in Figs.~\ref{fig:PinFin_Row5_NusseltPin}(b,c) indicate that relative to the baseline EV model, the SPWB correction can effectively decrease \(\mathrm{Nu}\) on upwind surface to a level lower than the LES result, whereas the DSDL model can effectively increase \(\mathrm{Nu}\) on partial downwind surface to a level higher than the LES result. When the DSDL model with \(\beta=0\sim1\) is combined with the SPWB correction with \(\alpha=0.5\), we obtain the resulting \(\mathrm{Nu}\) distributions shown in Fig.~\ref{fig:PinFin_Row5_NusseltPin}(d). Predictions of the combined model SPWB-DSDL well agree with the LES data on upwind surface with \(\beta_\mathrm{pin}=0^\circ\sim60^\circ\) and on downwind surface with \(\beta_\mathrm{pin}=135^\circ\sim180^\circ\). Errors of the three simulations are \(0\%\sim4\%\) at stagnation point and are \(1\%\sim7\%\) at \(\beta_\mathrm{pin}=180^\circ\). A noticeable discrepancy between the SPWB-DSDL results and the LES data is present near the separation point, in the range of \(\beta_\mathrm{pin}=105^\circ\sim125^\circ\), where the errors are up to \(40\%\sim50\%\). This underestimation of \(\mathrm{Nu}\) is closely related to the underestimation of \(k\) in the same region illustrated in Fig.~\ref{fig:PinFin_Row5_ktot_stream}.

\begin{figure}[htbp]
    \centering
    \includegraphics[width=1.0\linewidth]{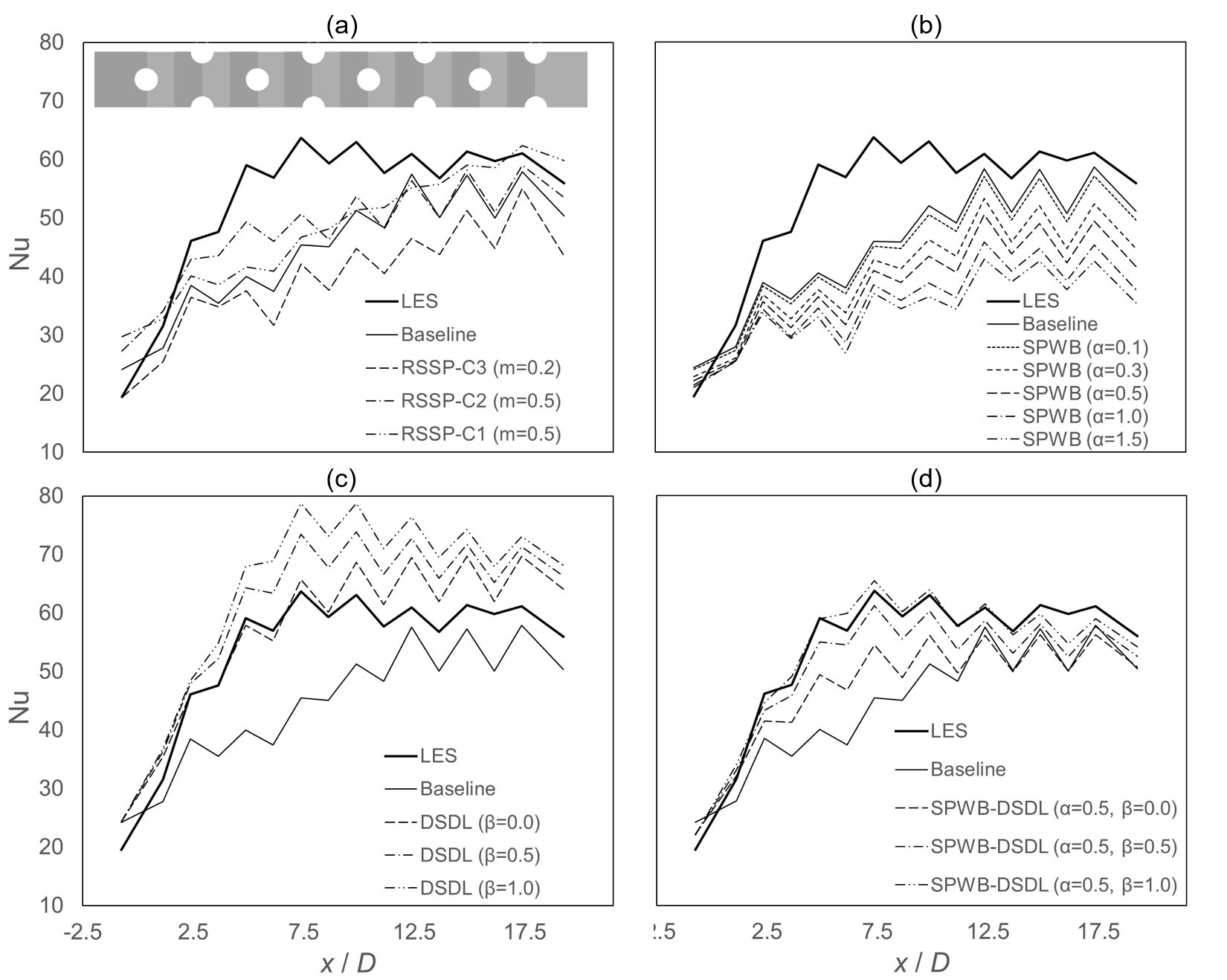}
    \caption{Distributions of Nusselt numbers (locally averaged on small patches) on the surface of fin}
    \label{fig:PinFin_NusseltFin}
\end{figure}

Lastly, we briefly examine the heat transfer on the fin surface. Divide the entire fin surface into 16 patches as illustrated in Fig.~\ref{fig:PinFin_NusseltFin}(a). Define the locally averaged Nusselt number \(\mathrm{Nu}\) for each patch, which is similar to Eq.~(\ref{eq:NusseltDef}) except that \((\partial_n\mathit{\Theta})_w\) is the average wall-normal \(\mathit{\Theta}\) gradient on each fin patch and \(\mathit{\Theta}_b\) is the flow-rate-weighted average \(\mathit{\Theta}\) in the space over each patch. Fig.~\ref{fig:PinFin_NusseltFin} shows the distributions of \(\mathrm{Nu}\). In Fig.~\ref{fig:PinFin_NusseltFin}(a), the baseline and the RSSP simulations fail to predict a rapid increase of \(\mathrm{Nu}\) in the range of the first three pin rows. In Figs.~\ref{fig:PinFin_NusseltFin}(b,c), the simulations of SPWB and DSDL respectively underpredict and overpredict \(\mathrm{Nu}\) on almost the entire fin surface. In general, the best predictions result from the simulations SPWB-DSDL(\(\alpha\)=0.5, \(\beta\)=0.5) and SPWB-DSDL(\(\alpha\)=0.5, \(\beta\)=1.0), as shown in Figs.~\ref{fig:PinFin_NusseltFin}(d). They also have the smallest absolute root-mean-square errors \(\Delta\mathrm{Nu}\) among all the simulations, as shown in Tab.~\ref{tab:PinFin_NusseltFinError}.


\begin{table}
\footnotesize
\centering
\label{tab:PinFin_NusseltFinError}
\begin{tabular}{p{0.3\linewidth}<{\centering} p{0.3\linewidth}<{\centering} p{0.3\linewidth}<{\centering}}
\toprule
Simulation Category & Parameter(s) & Absolute Error \(\Delta\mathrm{Nu}\) \\
\midrule
Baseline & N/A & 11.1 \\
\hline
\multirow{3}*{RSSP-C\(n\)} & \(n=3\), \(m=0.2\) & 15.5 \\
                           & \(n=2\), \(m=0.5\) &  7.8 \\
                           & \(n=1\), \(m=0.5\) &  9.4 \\
\hline
\multirow{3}*{SPWB} & \(\alpha=0.1\) & 11.4 \\
                    & \(\alpha=0.3\) & 14.3 \\
                    & \(\alpha=0.5\) & 16.3 \\
                    & \(\alpha=1.0\) & 19.2 \\
                    & \(\alpha=1.5\) & 20.8 \\
\hline
\multirow{3}*{DSDL} & \(\beta=0.0\) &  5.0 \\
                    & \(\beta=0.5\) &  8.2 \\
                    & \(\beta=1.0\) & 11.4 \\
\hline
\multirow{3}*{SPWB-DSDL} & \(\alpha=0.5\), \(\beta=0.0\) & 7.2 \\
                         & \(\alpha=0.5\), \(\beta=0.5\) & 3.5 \\
                         & \(\alpha=0.5\), \(\beta=1.0\) & 2.0 \\
\bottomrule
\end{tabular}
\caption{Absolute root-mean-square errors of locally averaged Nusselt numbers on the fin}
\end{table}

\subsubsection{Summary}
\label{Subsubsec:PinFinSummary}

In this subsection, we implement five categories of steady-RANS simulations for flow and heat transfer in a pin-fin array. The case features an upstream impingement flow pattern and a downstream quasi-two-dimensional separation flow pattern around one of the cylindrical pins. All simulation results are compared against the reference data in \cite{Hao19}.

The baseline EV model significantly overpredicts the size of the recirculation region and underpredicts the turbulence kinetic energy level downstream of the pin; it also tends to substantially overpredict wall-normal Reynolds stress and heat transfer on the upwind surface while underpredict the same two properties on the downwind surface. The RSSP method can correct the baseline model's prediction of the recirculation region size, but is apparently incapable of fixing any of the other above-mentioned problems. By contrast, the SPWB-DSDL method is essentially more capable of fixing all of the above-mentioned problems of the baseline model. The ultimate results of heat transfer predicted by SPWB-DSDL(\(\alpha\)=0.5, \(\beta\)=0.5\(\sim\)1.0) fairly agree with the reference data on most of the pin surface and the fin surface.

One remaining problem to be fixed is on the region near separation onset point, where energy and thus heat transfer are still apparently underestimated. In the future, a much stronger CS production formulation without increasing the shear stress could be used near the separation point to address this problem.

\subsection{Scalar dispersion around a skewed bump}
\label{Subsec:SkewedBump}

\subsubsection{Flow configuration, reference database, and simulation set-ups}
\label{Subsubsec:BumpConfiguration}

The second test case is scalar dispersion around a skewed bump mounted on a wall. The case features an upstream concave surface flow and a downstream complicated three-dimensional separation. The flow dynamics of this case was first studied using experiment by \cite{Ching18a,Ching18b} and LES by \cite{Ching20}, primarily focusing on the dynamics of three-dimensional separation bubble and unsteady vortex structures. In the present work, we selected the bump case with the skew angle of \(10^\circ\) in \cite{Ching20}, introduced two independent passive scalars to the flow field, and implemented a new LES for turbulent transport of the two scalars. The results of this LES and of the previous work \cite{Ching20} will be used as the reference data in this subsection.

\begin{figure}[htbp]
    \centering
    \includegraphics[width=0.95\linewidth]{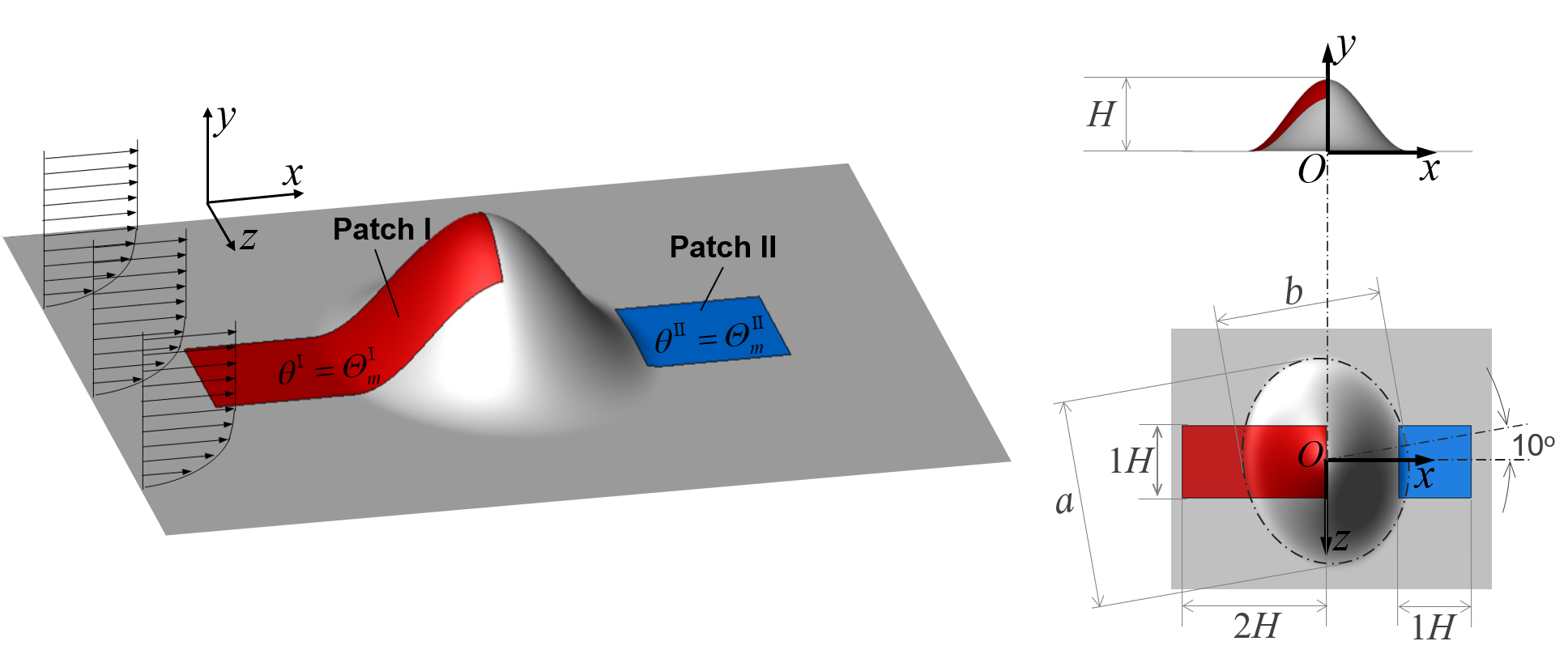}
    \caption{Configuration of the flow past a skewed bump with patches of surface sources of two independent passive scalars}
    \label{fig:Bump_Configuration}
\end{figure}

Fig.~\ref{fig:Bump_Configuration} shows the flow configuration for this case. A \(10^\circ\) skewed bump with a height \(H\) is installed on a flat plate with \(y=0\). The bump has cosine cross-sections viewed from the side and elliptical cross-sections viewed from above. On the plane \(y=0\), the cross-section has a major axis \(a=3H\) and a minor axis \(b=9H/4\). The coordinate origin \(O\) is set at the bump center. More details on the bump geometry can be found in \cite{Ching20}. The incoming flow is in \(x\)-direction, and has a boundary layer profile with a momentum thickness \(0.042H\) and a shape factor 1.7 at \(x=-4H\). The Reynolds number \(\mathrm{Re}=U_bH/\nu=16,000\) where \(U_b\) is the bulk velocity. A \(2H\times1H\) patch of surface source of scalar \(\theta^\mathrm{I}\) is placed partially on the upwind surface of the bump and partially on the flat plate upstream. \(\theta^\mathrm{I}\) equals to a constant \(\mathit{\Theta}^\mathrm{I}_m\) on this patch and is adiabatic on any piece of solid surface outside of this patch. Similarly, a smaller \(1H\times1H\) patch of surface source of scalar \(\theta^\mathrm{II}\) is placed downstream of the bump. \(\theta^\mathrm{II}\) equals to a constant \(\mathit{\Theta}^\mathrm{II}_m\) on this patch and is adiabatic on any piece of solid surface outside of this patch. In this problem, \(\theta^\mathrm{I}\) and \(\theta^\mathrm{II}\) are passive scalars and transported independently of each other. At inlet, both \(\theta^\mathrm{I}\) and \(\theta^\mathrm{II}\) are zero. The molecular Schmidt numbers \(\mathrm{Sc}=\nu/\kappa\) are \(1\) for both \(\theta^\mathrm{I}\) and \(\theta^\mathrm{II}\).

For LES of this scalar dispersion problem, the present work directly employed the set-ups provided in \cite{Ching20}. For steady-RANS simulations of this problem, the inlet boundary is set at \(x=-6H\), where the conditions of profiles of mean velocity, turbulence kinetic energy, and turbulence length scale are set to be consistent with the LES data. The number of cells used to discretize the RANS computational domain is around 5.81 million.

The discussions of steady-RANS results in this subsection will focus on the two categories: RSSP and SPWB-DSDL. The parameter \(m\) values for RSSP-C1, RSSP-C2 and RSSP-C3 are 0.5, 0.5 and 0.1, respectively. The parameter \(\alpha\) for SPWB-DSDL is 0.5, which is same with that used for the pin-fin array case in \S\ref{Subsec:PinFinArray}.

Compared to the bluff bodies in \S\ref{Subsec:PinFinArray} and \cite{Hao21}, the bump configuration in this subsection is a much more streamlined obstacle. Our preparatory tests demonstrated that for the SPWB-DSDL model in this case, using a \(c_{tr}=1.5\sim1.8\) as suggested by \cite{Hao21} for bluff body flows could result in total disappearance of mean flow separation and significant overprediction of downstream energy. This nonphysical result implied that for flows past streamlined obstacles as this bump, one should consider using larger \(c_{tr}\). Therefore in this subsection, we will examine the effect of various \(c_{tr}\) in a larger range while ignore the scale separation effect, i.e. \(\beta=0\), for the SPWB-DSDL model.

\subsubsection{Results of mean flow}
\label{Subsubsec:BumpResultsMeanFlow}

\begin{table}
\footnotesize
\centering
\label{tab:Bump_Bubble}
\begin{tabular}{p{0.3\linewidth}<{\centering} p{0.3\linewidth}<{\centering} p{0.3\linewidth}<{\centering}}
\toprule
Simulation Category & Parameter(s) & Bubble Size \(X_m/H\) \\
\midrule
LES      & N/A & 2.41 \\
\hline
Baseline & N/A & 2.97 \\
\hline
\multirow{3}*{RSSP-C\(n\)} & \(n=3\), \(m=0.1\) & 3.28 \\
                           & \(n=2\), \(m=0.5\) & 2.16 \\
                           & \(n=1\), \(m=0.5\) & 1.51 \\
\hline
\multirow{3}*{SPWB-DSDL} & \(\alpha=0.5\), \(c_{tr}=7.0\) & 2.46 \\
                         & \(\alpha=0.5\), \(c_{tr}=5.0\) & 2.29 \\
                         & \(\alpha=0.5\), \(c_{tr}=3.0\) & 1.75 \\
\bottomrule
\end{tabular}
\caption{Size of separation bubble downstream of the bump}
\end{table}

\begin{figure}[htbp]
    \centering
    \includegraphics[width=1.0\linewidth]{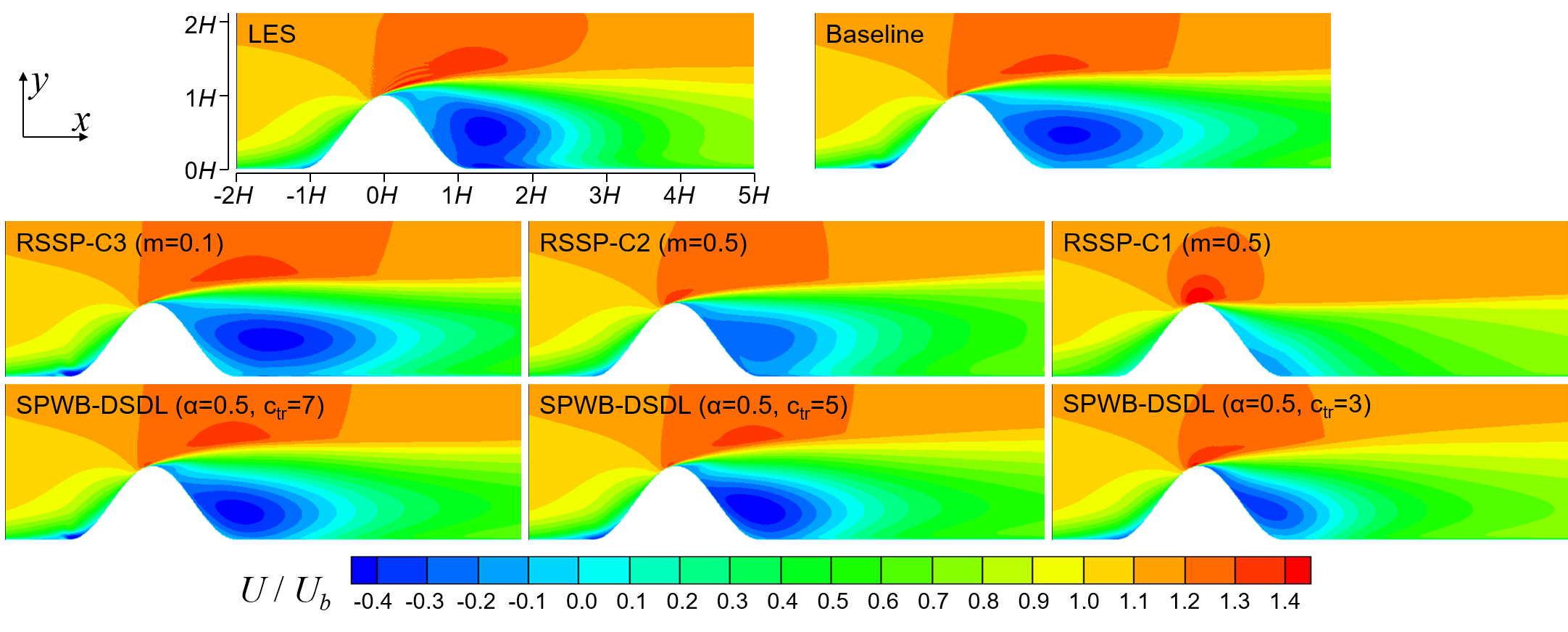}
    \caption{Contours of streamwise mean velocity \(U\) on the plane \(z=0\)}
    \label{fig:Bump_z=0_Ux}
\end{figure}

\begin{figure}[htbp]
    \centering
    \includegraphics[width=1.0\linewidth]{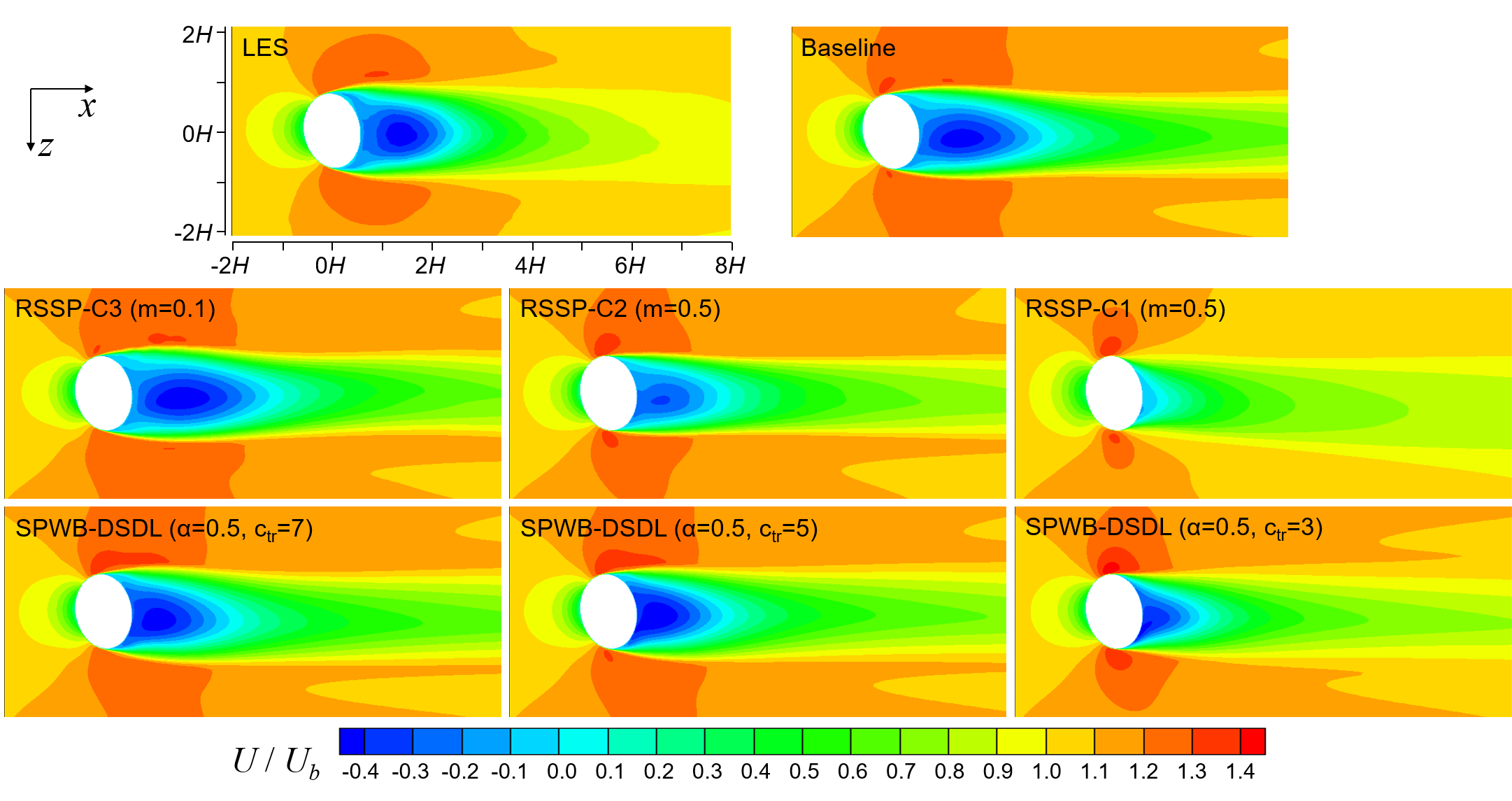}
    \caption{Contours of streamwise mean velocity \(U\) on the plane \(y=0.5H\)}
    \label{fig:Bump_y=0.5_Ux}
\end{figure}

The contours of \(x\)-mean velocity \(U\) in Figs.~\ref{fig:Bump_z=0_Ux} and \ref{fig:Bump_y=0.5_Ux} illustrate the separation bubble behind the bump. Tab.~\ref{tab:Bump_Bubble} lists the bubble size \(X_m\) resulting from each simulation, which is defined as the maximum \(x\)-coordinate among all locations on the \(z=0\) plane where \(U<0\). Compared to the LES result, the baseline EV model overpredicts the bubble size \(X_m\) by 34\%. This overprediction can be effectively corrected by RSSP-C2 and RSSP-C1, whose results of \(X_m\) are respectively 15\% and 54\% smaller than the LES result. The SPWB-DSDL model is also capable of correcting the baseline model's overprediction of bubble size. The \(X_m\) values resulting from the three simulations with \(c_{tr}\)=7.0, 5.0, and 3.0 are respectively 3\% larger, 7\% smaller, and 39\% smaller than the LES result.

\subsubsection{Results of turbulence energy and stress}
\label{Subsubsec:BumpResultsTurbulence}

\begin{figure}[htbp]
    \centering
    \includegraphics[width=1.0\linewidth]{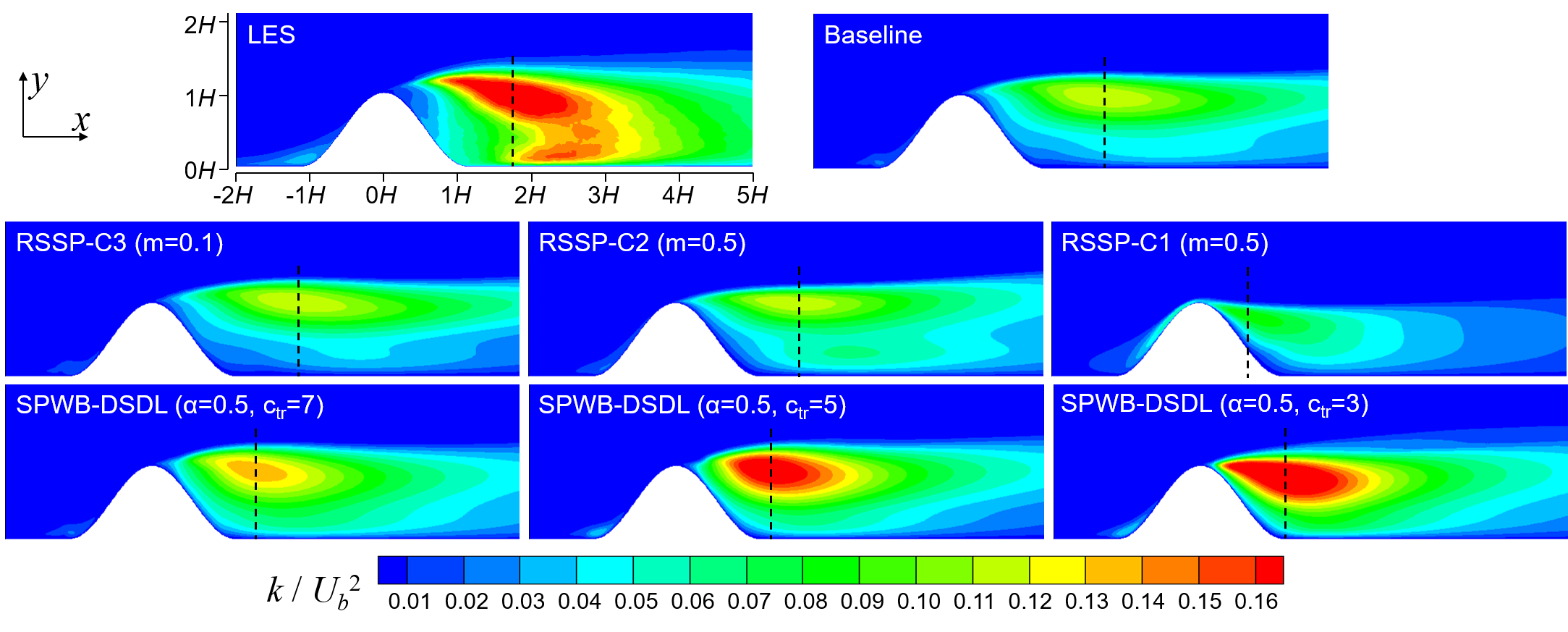}
    \caption{Contours of turbulence kinetic energy \(k\) on the plane \(z=0\). The \(y\)-directional dashed line in each panel is through the location with the maximum \(k\) on the \(z=0\) plane.}
    \label{fig:Bump_z=0_k}
\end{figure}

The contours of turbulence kinetic energy \(k\) in Fig.~\ref{fig:Bump_z=0_k} illustrate the turbulence intensity level in the bubble-induced free shear layer on the \(z=0\) plane. Qualitatively, the baseline model substantially underestimates the turbulence intensity in the shear layer, and none of the three RSSP simulations is capable of correcting this underestimation. By contrast, the shear-layer turbulence intensity resulting from the SPWB-DSDL model with \(c_{tr}\)=5.0 and 3.0 is fairly comparable with the LES data, although the SPWB-DSDL predicted location of the high intensity region is \(0.4H\sim0.6H\) ahead of that of LES result.

\begin{figure}[htbp]
    \centering
    \includegraphics[width=1.0\linewidth]{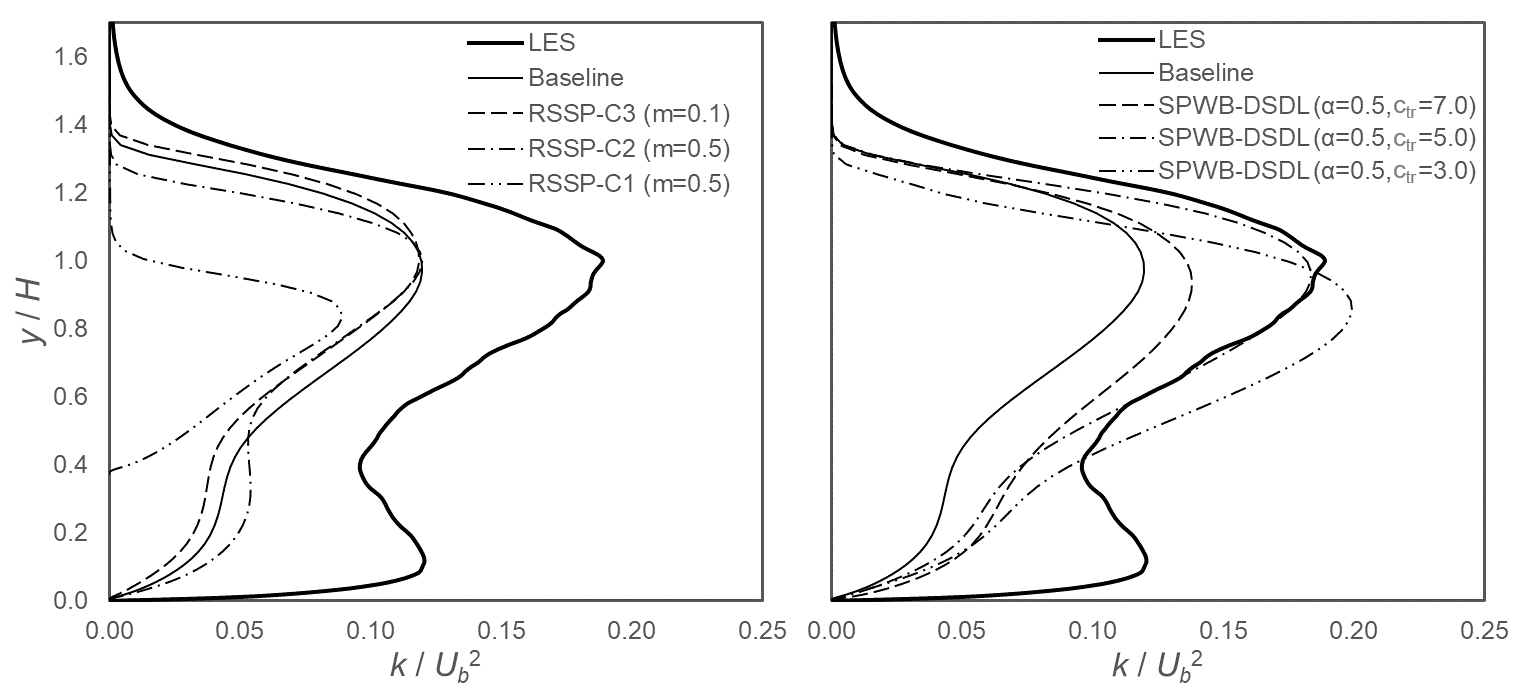}
    \caption{Profiles of turbulence kinetic energy \(k\) downstream of bump. The sample lines are indicated by the dashed lines in Fig.~\ref{fig:Bump_z=0_k}.}
    \label{fig:Bump_yProfile_k}
\end{figure}

\begin{figure}[htbp]
    \centering
    \includegraphics[width=1.0\linewidth]{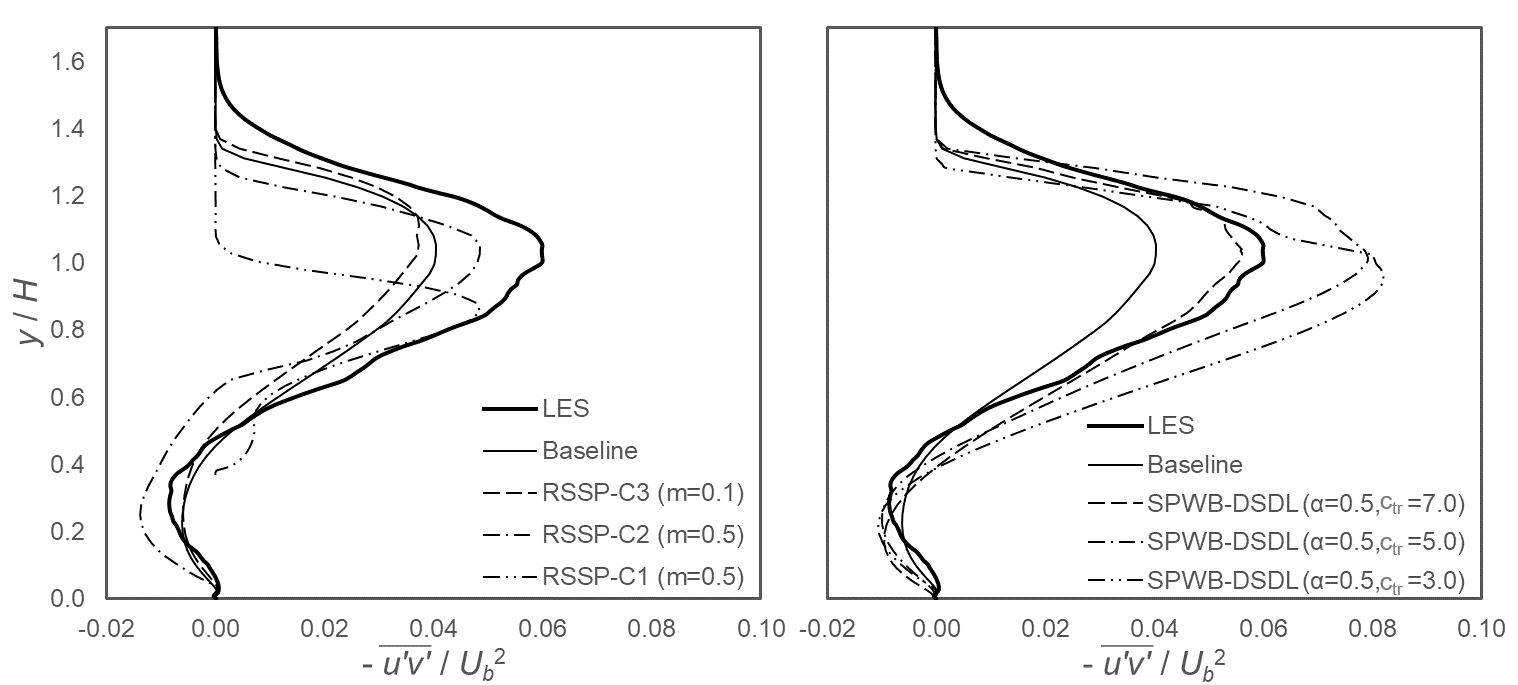}
    \caption{Profiles of turbulence shear stress \(\overline{u^\prime v^\prime}\) downstream of bump. The sample lines are indicated by the dashed lines in Fig.~\ref{fig:Bump_z=0_k}.}
    \label{fig:Bump_yProfile_Ruv}
\end{figure}

Quantitatively, in each panel of Fig.~\ref{fig:Bump_z=0_k} we select the \(y\)-directional sample line through the location with the maximum \(k\), and plot all the profiles of energy \(k\) and shear stress \(\overline{u^\prime v^\prime}\) in Figs.~\ref{fig:Bump_yProfile_k} and \ref{fig:Bump_yProfile_Ruv}. The figures show that the underprediction of \(k\) and \(\overline{u^\prime v^\prime}\) magnitudes by baseline and RSSP can be largely offset by SPWB-DSDL, except for \(k\) in the region \(y=0\sim0.4H\), which is to be discussed in \S\ref{Subsubsec:BumpResultsScalarRelease}. The errors of peak \(k\) values relative to LES are \(-53\%\sim-37\%\) for baseline and RSSP, \(-27\%\) for SPWB-DSDL(\(\alpha\)=0.5, \(c_{tr}\)=7.0), and \(-3\%\sim+5\%\) for SPWB-DSDL(\(\alpha\)=0.5, \(c_{tr}\)=5.0\(\sim\)3.0).
The errors of peak \(\overline{u^\prime v^\prime}\) magnitudes are \(-38\%\sim-19\%\) for baseline and RSSP, \(-5\%\) for SPWB-DSDL(\(\alpha\)=0.5, \(c_{tr}\)=7.0), and \(+33\%\sim+37\%\) for SPWB-DSDL(\(\alpha\)=0.5, \(c_{tr}\)=5.0\(\sim\)3.0).

\begin{figure}[htbp]
    \centering
    \includegraphics[width=1.0\linewidth]{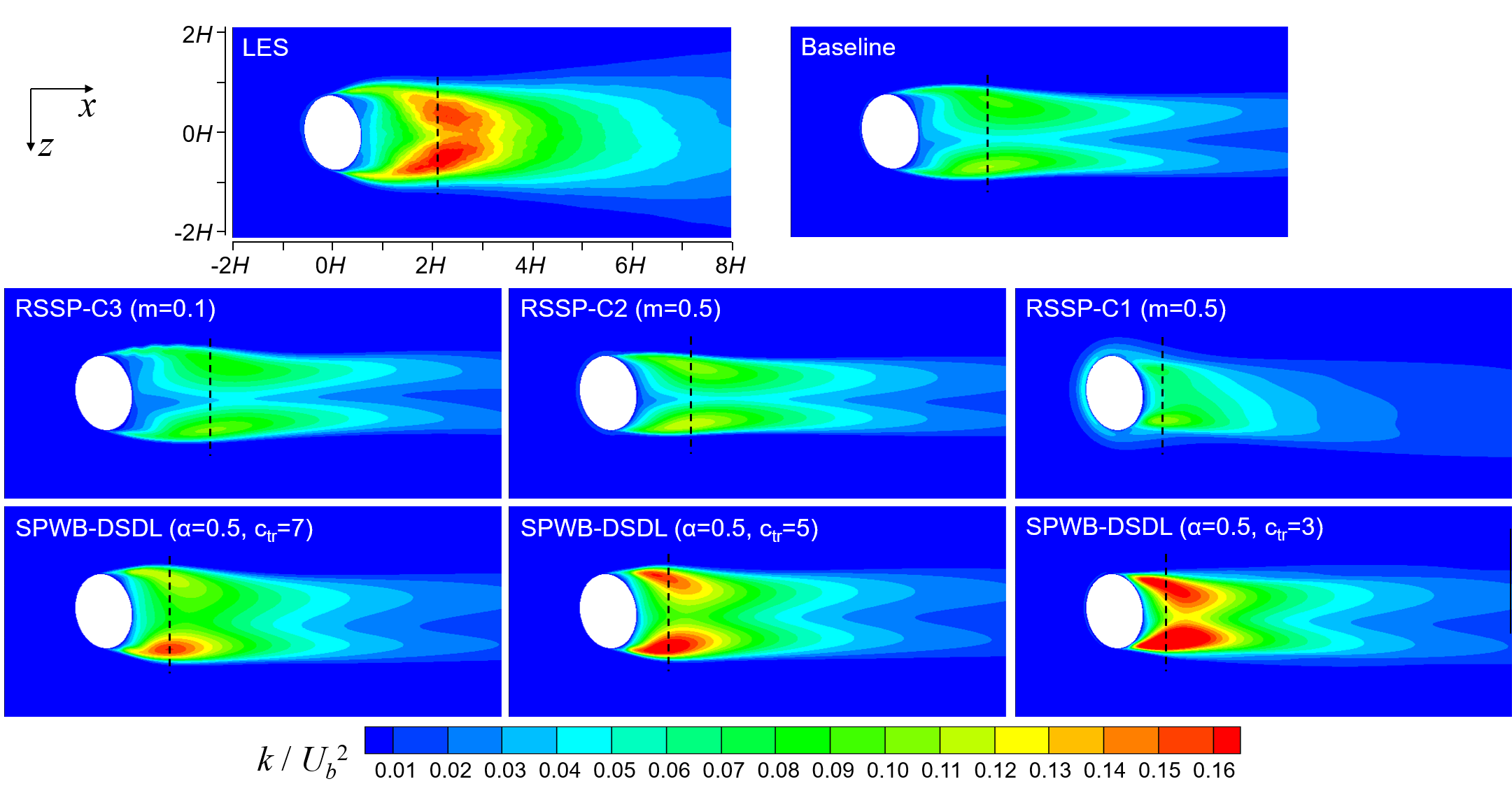}
    \caption{Contours of turbulence kinetic energy \(k\) on the plane \(y=0.5H\). The \(z\)-directional dashed line in each panel is through the location with the maximum \(k\) on the \(y=0.5H\) plane.}
    \label{fig:Bump_y=0.5_k}
\end{figure}

\begin{figure}[htbp]
    \centering
    \includegraphics[width=1.0\linewidth]{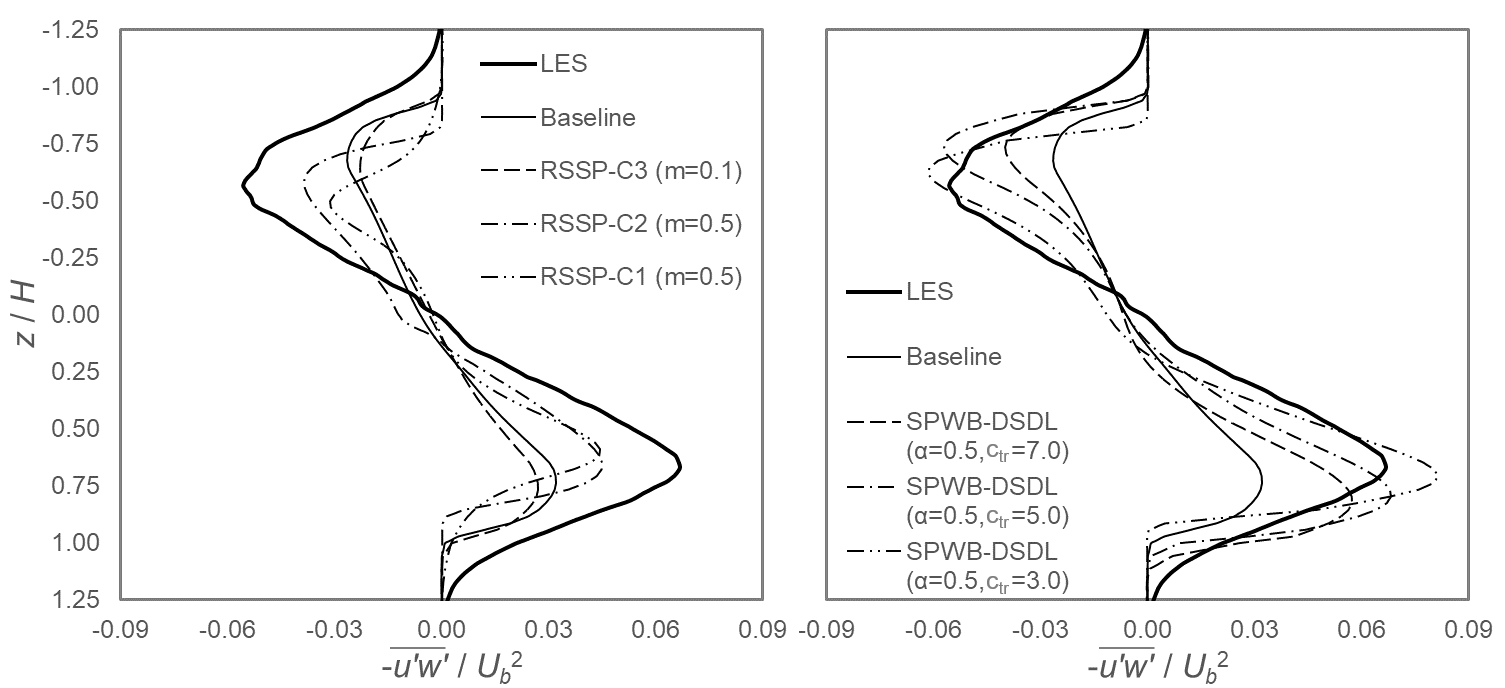}
    \caption{Profiles of turbulence shear stress \(\overline{u^\prime w^\prime}\) downstream of bump. The sample lines are indicated by the dashed lines in Fig.~\ref{fig:Bump_y=0.5_k}.}
    \label{fig:Bump_zProfile_Ruw}
\end{figure}

The turbulence kinetic energy and shear stress in the bubble-induced free shear layer on the \(y=0.5H\) plane are illustrated in Figs.~\ref{fig:Bump_y=0.5_k}\(\sim\)\ref{fig:Bump_zProfile_Ruw}. All simulations exhibit fairly similar behaviors to those for the free shear layer on the \(z=0\) plane illustrated in Figs.~\ref{fig:Bump_z=0_k}\(\sim\)\ref{fig:Bump_yProfile_Ruv}.
\begin{figure}[htbp]
    \centering
    \includegraphics[width=1.0\linewidth]{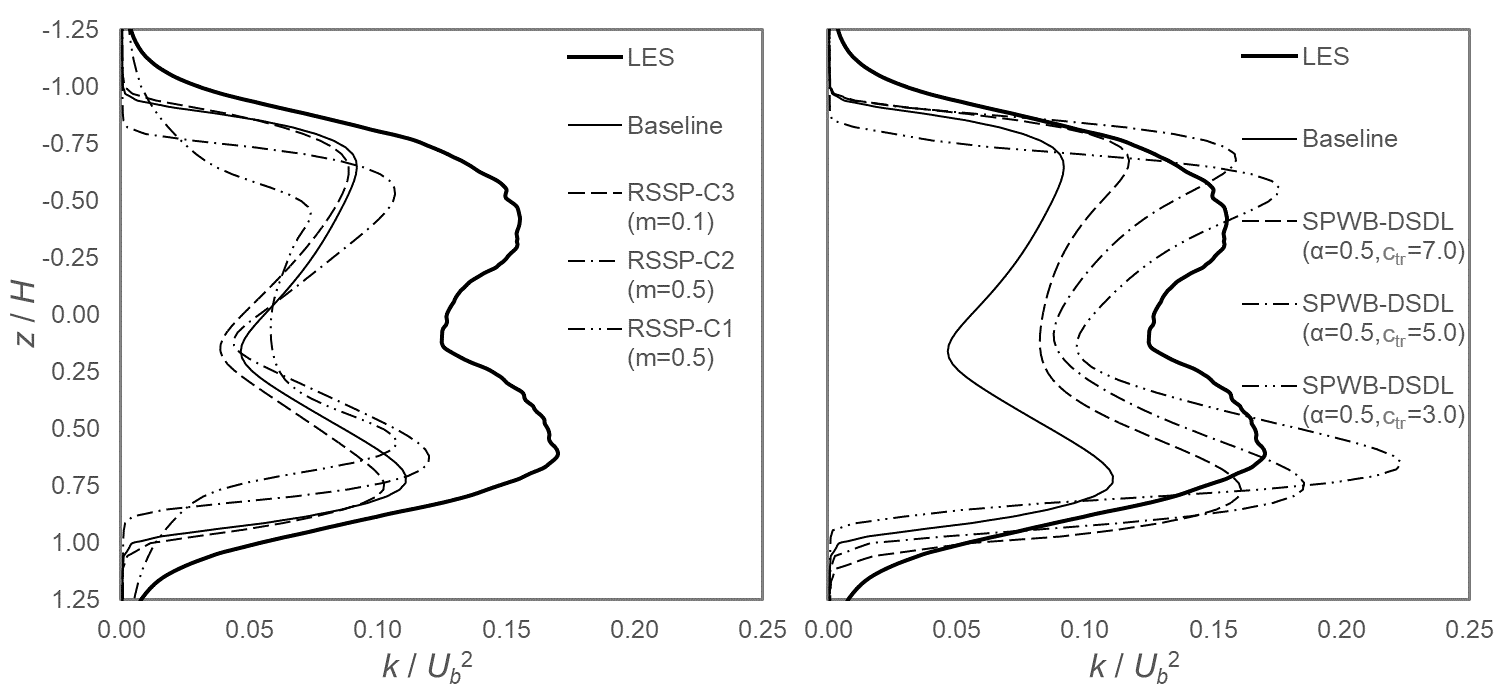}
    \caption{Profiles of turbulence kinetic energy \(k\) downstream of bump. The sample lines are indicated by the dashed lines in Fig.~\ref{fig:Bump_y=0.5_k}.}
    \label{fig:Bump_zProfile_k}
\end{figure}

\begin{figure}[htbp]
    \centering
    \includegraphics[width=1.0\linewidth]{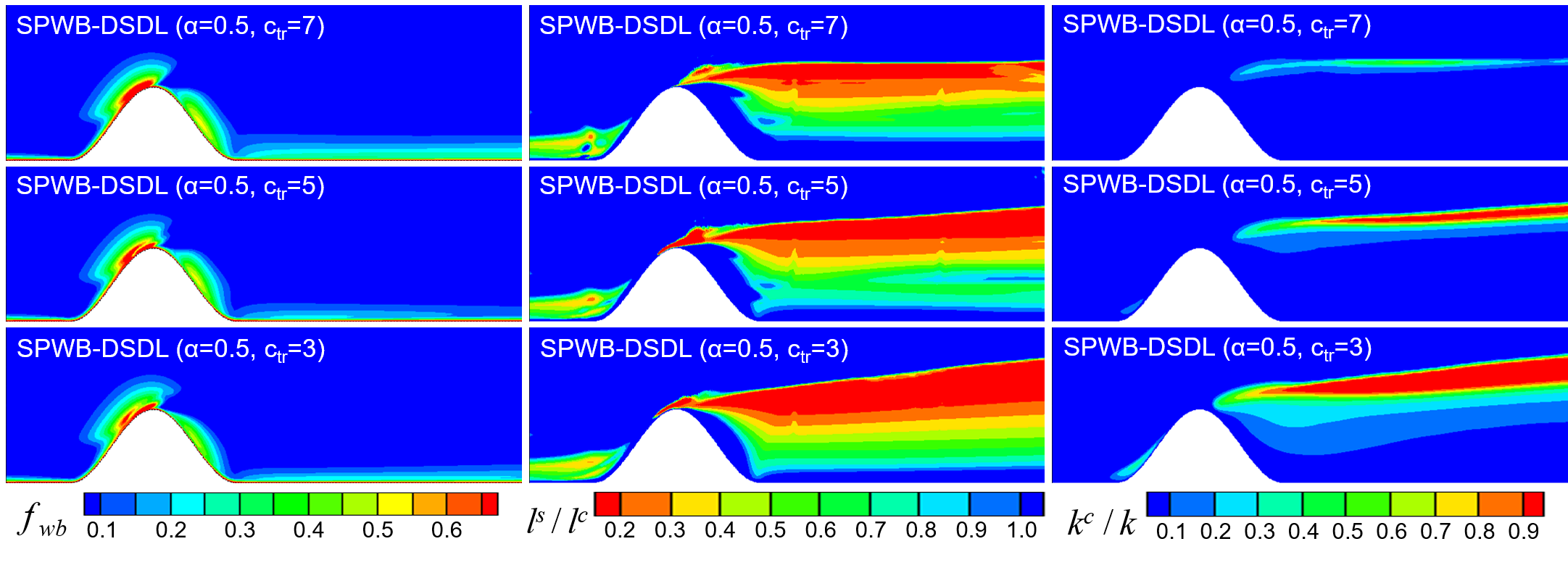}
    \caption{Contours of wall blocking function \(f_{wb}\) (left column), length scalar ratio \(l^s/l^c\) (middle column), and CS energy fraction \(k^c/k\) (right column) for three SPWB-DSDL simulations respectively with \(c_{tr}=7.0\) (upper row), \(5.0\) (middle row) and \(3.0\) (bottom row) on the plane \(z=0\)}
    \label{fig:Bump_z=0_Blending}
\end{figure}
\begin{figure}[htbp]
    \centering
    \includegraphics[width=1.0\linewidth]{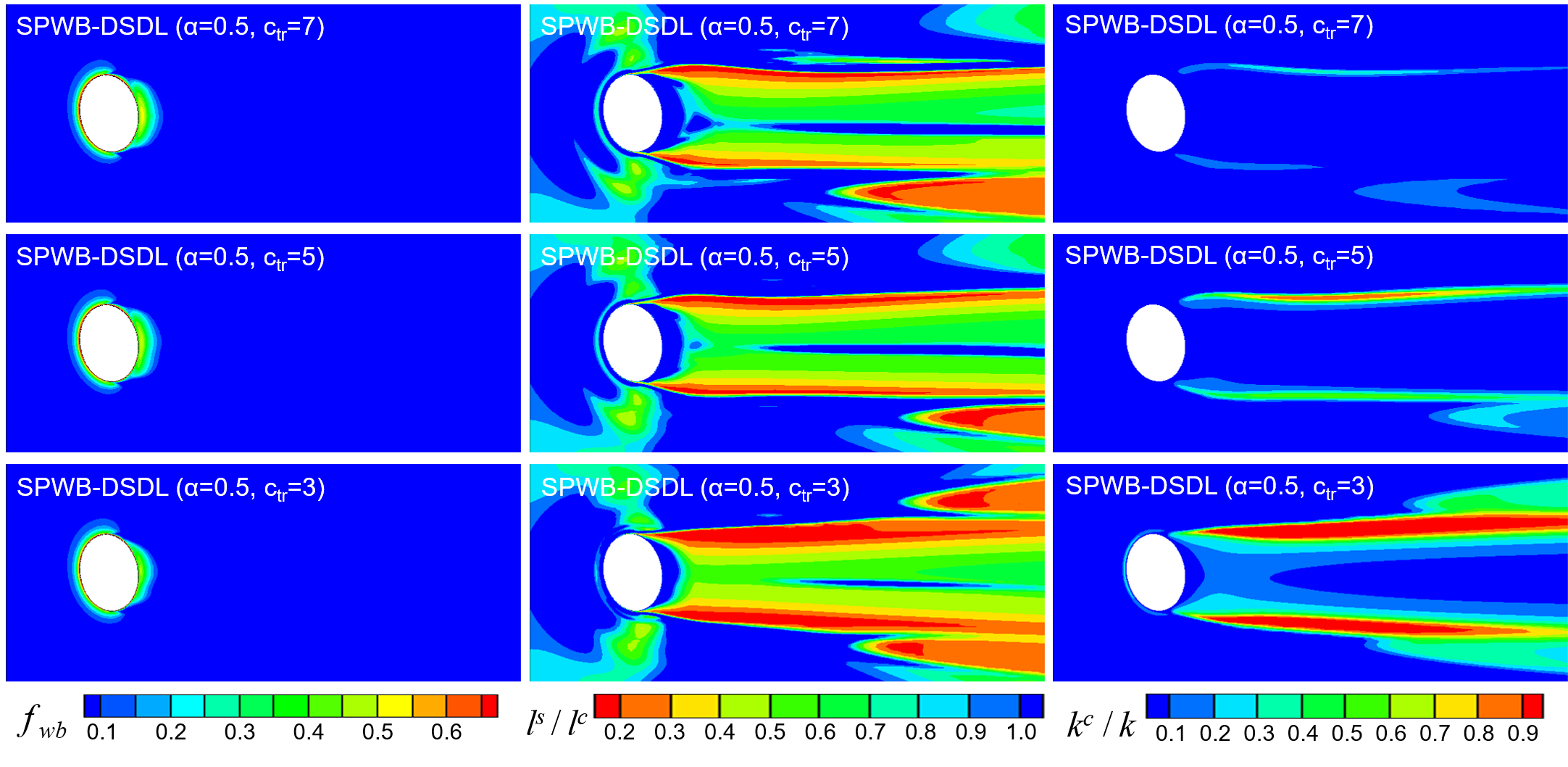}
    \caption{Contours of wall blocking function \(f_{wb}\) (left column), length scalar ratio \(l^s/l^c\) (middle column), and CS energy fraction \(k^c/k\) (right column) for three SPWB-DSDL simulations respectively with \(c_{tr}=7.0\) (upper row), \(5.0\) (middle row) and \(3.0\) (bottom row) on the plane \(y=0.5H\)}
    \label{fig:Bump_y=0.5_Blending}
\end{figure}

Figs.~\ref{fig:Bump_z=0_Blending} and \ref{fig:Bump_y=0.5_Blending} show the contours of wall-blocking function \(f_{wb}\), length scale ratio \(l^s/l^c\), and CS energy fraction \(k^c/k\) predicted by SPWB-DSDL on the planes \(z=0\) and \(y=0.5H\). The \(f_{wb}\) contours in Fig.~\ref{fig:Bump_z=0_Blending} illustrate a high \(f_{wb}\) region covering over a half of the upwind surface. The region is crucial to avoiding the overestimation of scalar transport there, which will be seen in \S\ref{Subsubsec:BumpResultsScalarRelease}. In the length scale ratio \(l^s/l^c\) contours of the two figures, the free shear layers downstream of the bump are identified, where the energy fraction \(k^c/k\) increases as \(c_{tr}\) decreases from 7.0 to 3.0. These contours explain the high levels of energy and shear stress resulting from the SPWB-DSDL model, as observed in Figs.~\ref{fig:Bump_z=0_k}\(\sim\)\ref{fig:Bump_zProfile_Ruw}.

\begin{figure}[htbp]
    \centering
    \includegraphics[width=1.0\linewidth]{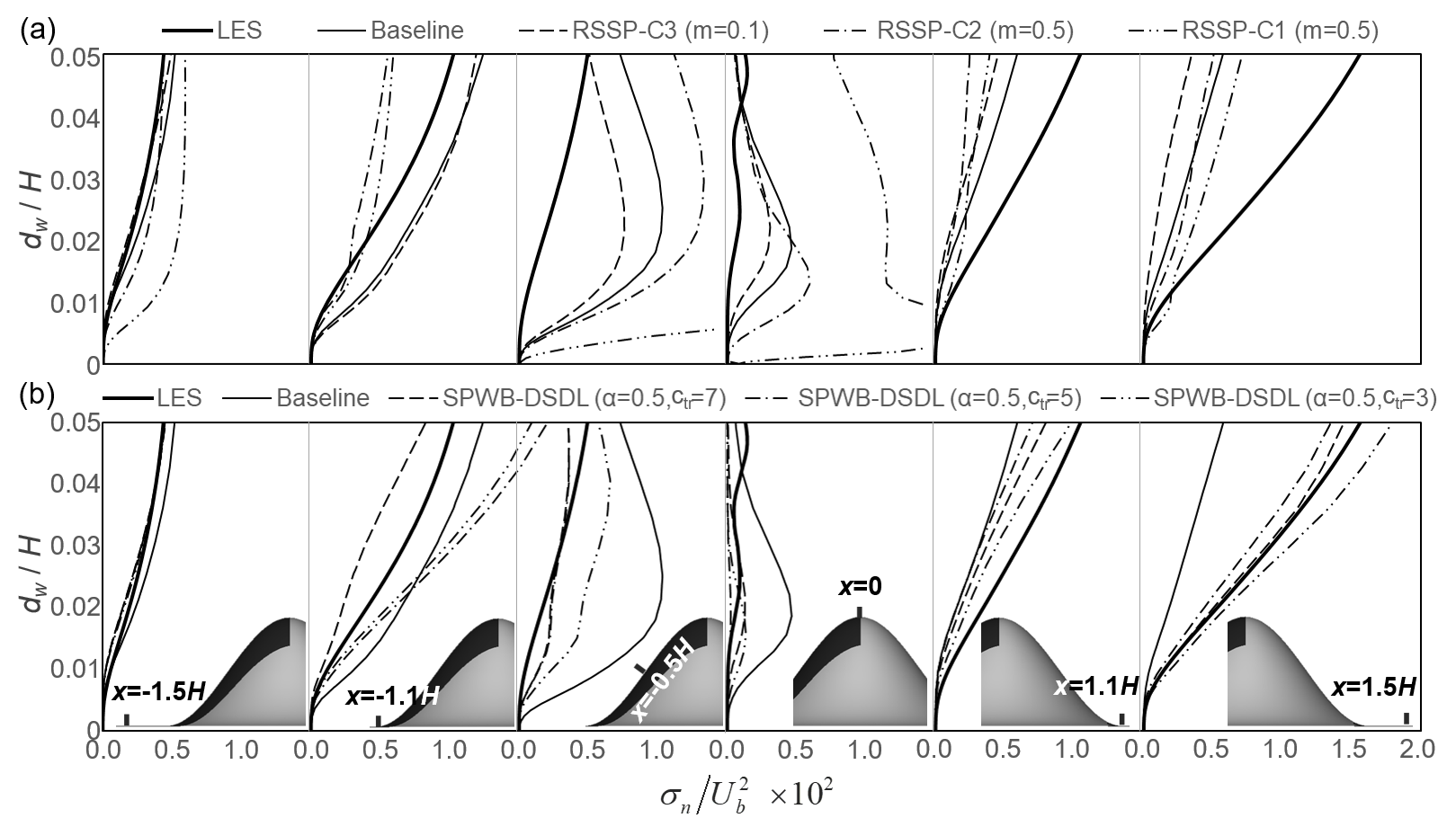}
    \caption{Near-wall profiles of wall-normal Reynolds stress \(\sigma_n\) versus wall distance \(d_w\) at six different locations (\(x/H=-1.5, -1.1, -0.5, 0, 1.1 \textup{ and } 1.5\) with \(z=0\) for all)}
    \label{fig:Bump_WNstress}
\end{figure}

Fig.~\ref{fig:Bump_WNstress} plots the near-wall profiles of wall-normal stress \(\sigma_n\) at six different locations on the surface. At \(x=-0.5H\) and \(0\) where the boundary layer is subjected to strong mean flow contraction in wall-normal direction, the baseline EV model considerably overpredicts the near-wall \(\sigma_n\) level. This overprediction cannot be effectively corrected by any of the RSSP simulations (Fig.~\ref{fig:Bump_WNstress}(a)), but is substantially alleviated by the SPWB-DSDL model (Fig.~\ref{fig:Bump_WNstress}(b)). By contrast, at \(x=1.1H\) and \(1.5H\) where the wall is covered by the separation bubble, the baseline model considerably underpredicts \(\sigma_n\). Again the performance of SPWB-DSDL is generally superior to that of RSSP in terms of correcting this underprediction.

\subsubsection{Results of scalar release on sources}
\label{Subsubsec:BumpResultsScalarRelease}

\begin{figure}[htbp]
    \centering
    \includegraphics[width=1.0\linewidth]{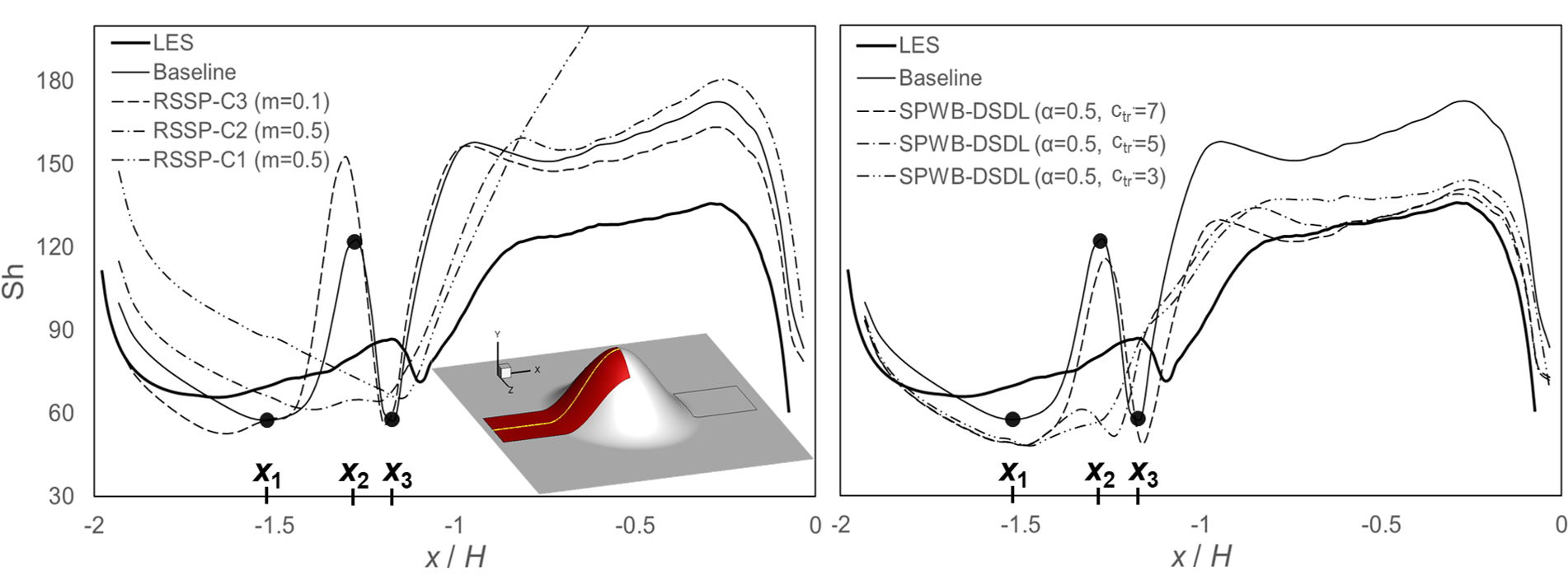}
    \caption{Distributions of Sherwood numbers \(\mathrm{Sh}\) along the centerline (\(z=0\)) of Patch I. The three solid circles marked on the \(\mathrm{Sh}\) curve are local extrema with \(x\)-coordinates denoted by \(x_1\), \(x_2\) and \(x_3\), respectively.}
    \label{fig:Bump_Sherwood1}
\end{figure}

Similar to Eq.~(\ref{eq:NusseltDef}), we define the local Sherwood number \(\mathrm{Sh}\) at a point on the surface source to quantify the local scalar transfer:
\begin{equation}
\label{eq:SherwoodDef}
    \mathrm{Sh}=-(\partial_n\mathit{\Theta})_w \, H \,/\, \mathit{\Theta}_m \,.
\end{equation}
Fig.~\ref{fig:Bump_Sherwood1} plots \(\mathrm{Sh}\) distributions along the centerlines, i.e.~\(z=0\), of the two source patches.

\begin{figure}[htbp]
    \centering
    \includegraphics[width=1.0\linewidth]{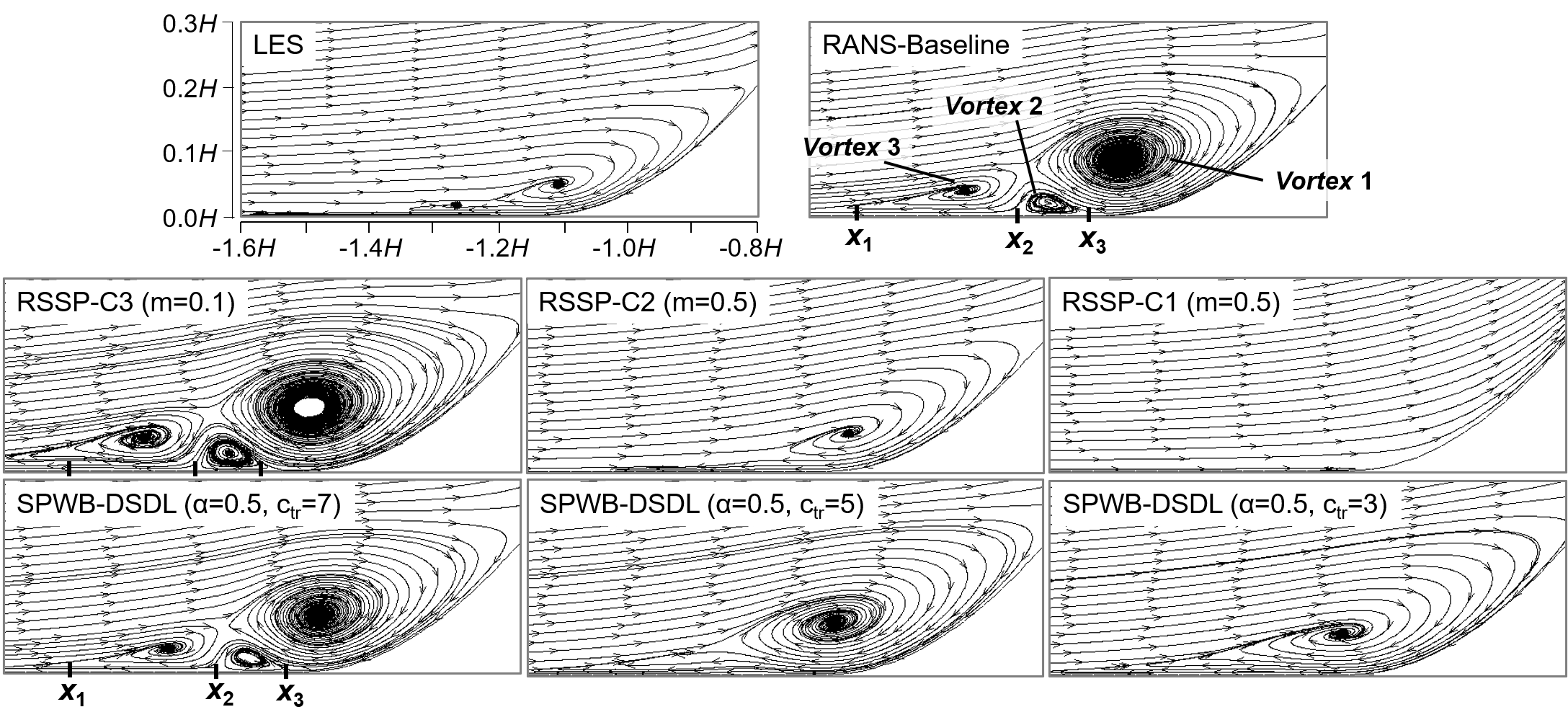}
    \caption{Mean streamlines on plane \(z=0\) near the upwind corner of bump. \(x_1\), \(x_2\) and \(x_3\) indicate the locations of \(\mathrm{Sh}\) extrema that are marked in Fig.~\ref{fig:Bump_Sherwood1}.}
    \label{fig:Bump_LeadingVortex}
\end{figure}

First consider the flat part of Patch I in the range of \(x=-2H\sim-1.1H\). A major discrepancy of \(\mathrm{Sh}\) between baseline model prediction and LES data is present at \(x=-1.5H\sim-1.1H\) where the former exhibits a striking increase followed by a dramatic decrease. Similar increase-decrease trend is also found in the results of RSSP-C3 and SPWB-DSDL(\(\alpha\)=0.5, \(c_{tr}\)=7.0). A more detailed examination in Fig.~\ref{fig:Bump_LeadingVortex} illustrates that this feature is associated with the three-vortex pattern (see \cite{Schwind62} for its formation mechanism) in the results of baseline model, RSSP-C3, and SPWB-DSDL(\(\alpha\)=0.5, \(c_{tr}\)=7.0) near upwind corner of the bump. The clockwise-rotating Vortex 3 in Fig.~\ref{fig:Bump_LeadingVortex} entrains the mainstream fluids with low \(\mathit{\Theta}\) into the boundary layer, thereby causing a large near-wall gradient of \(\mathit{\Theta}\) and the striking increase of \(\mathrm{Sh}\) from \(x_1\) to \(x_2\). Between Vortex 3 and the large main vortex, Vortex 1, is a small counterclockwise-rotating vortex, Vortex 2. It is a `backwater' with quite limited mean convection with the mainstream fluids, thus causing the dramatic decrease of \(\mathrm{Sh}\) from \(x_2\) to \(x_3\).

On the curved part of Patch I in the range of \(x=-0.8H\sim0\), the baseline model generally overpredicts \(\mathrm{Sh}\) by around 25\%. This number is slightly decreased to around 20\% by RSSP-C3, while even further increased by RSSP-C2 and RSSP-C1. The SPWB-DSDL model potently alleviates these overpredictions thanks to the model's wall-blocking mechanism illustrated by the \(f_{wb}\) contours in Fig.~\ref{fig:Bump_z=0_Blending}. The SPWB-DSDL(\(\alpha\)=0.5, \(c_{tr}\)=7.0\(\sim\)5.0) results fairly agree with the LES data, and the errors of SPWB-DSDL(\(\alpha\)=0.5, \(c_{tr}\)=3.0) results are lower than 10\%.

Between the above two ranges on Patch I, i.e.~in the range of \(x=-1.1H\sim-0.8H\), all the RANS simulations predict rapider increases of \(\mathrm{Sh}\) than that of LES data. This defect needs to be investigated in the future.

\begin{figure}[htbp]
    \centering
    \includegraphics[width=0.7\linewidth]{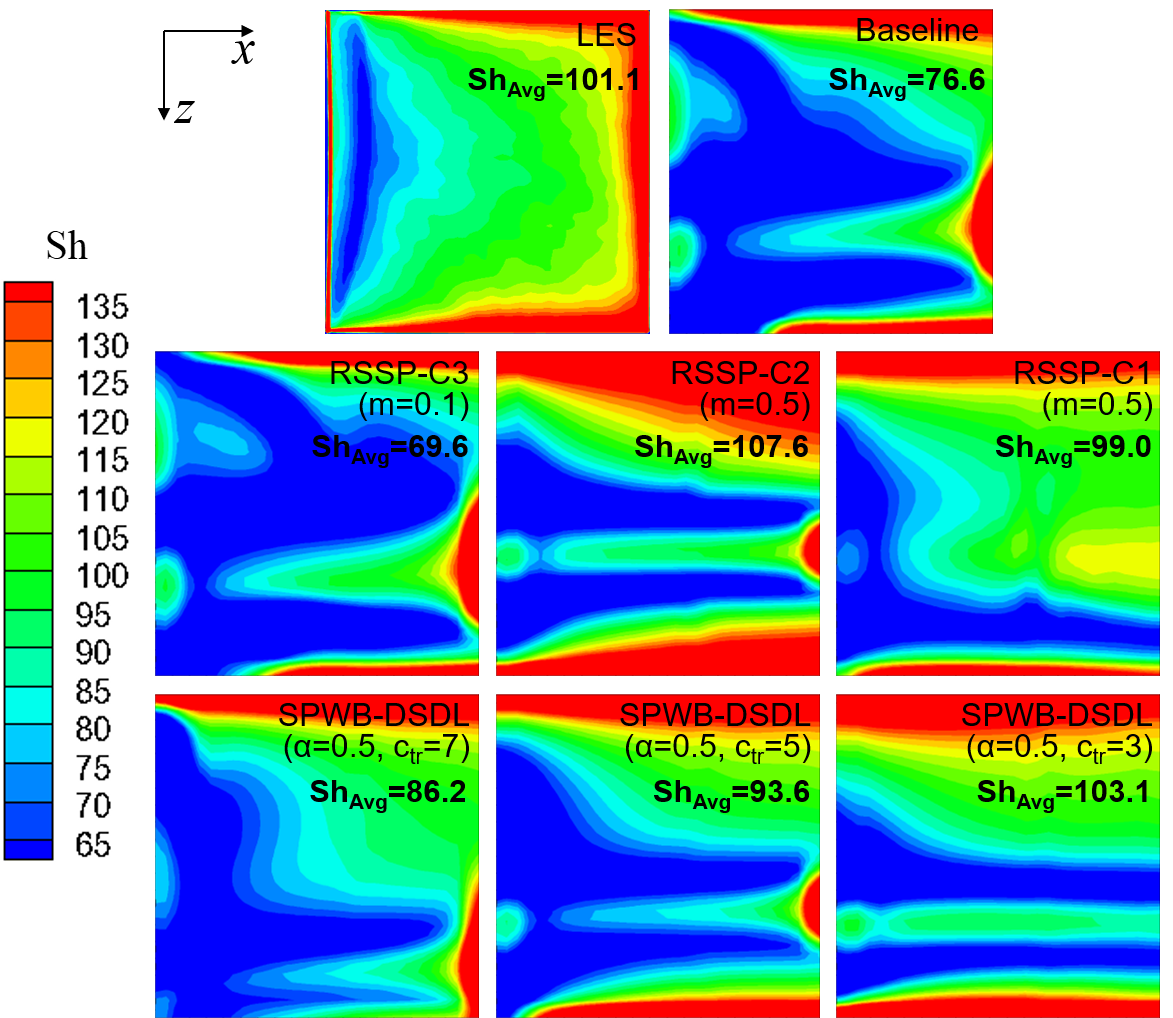}
    \caption{Contours of Sherwood numbers \(\mathrm{Sh}\) and their averages \(\mathrm{Sh_{Avg}}\) on Patch II}
    \label{fig:Bump_Sherwood2}
\end{figure}

\begin{figure}[htbp]
    \centering
    \includegraphics[width=1.0\linewidth]{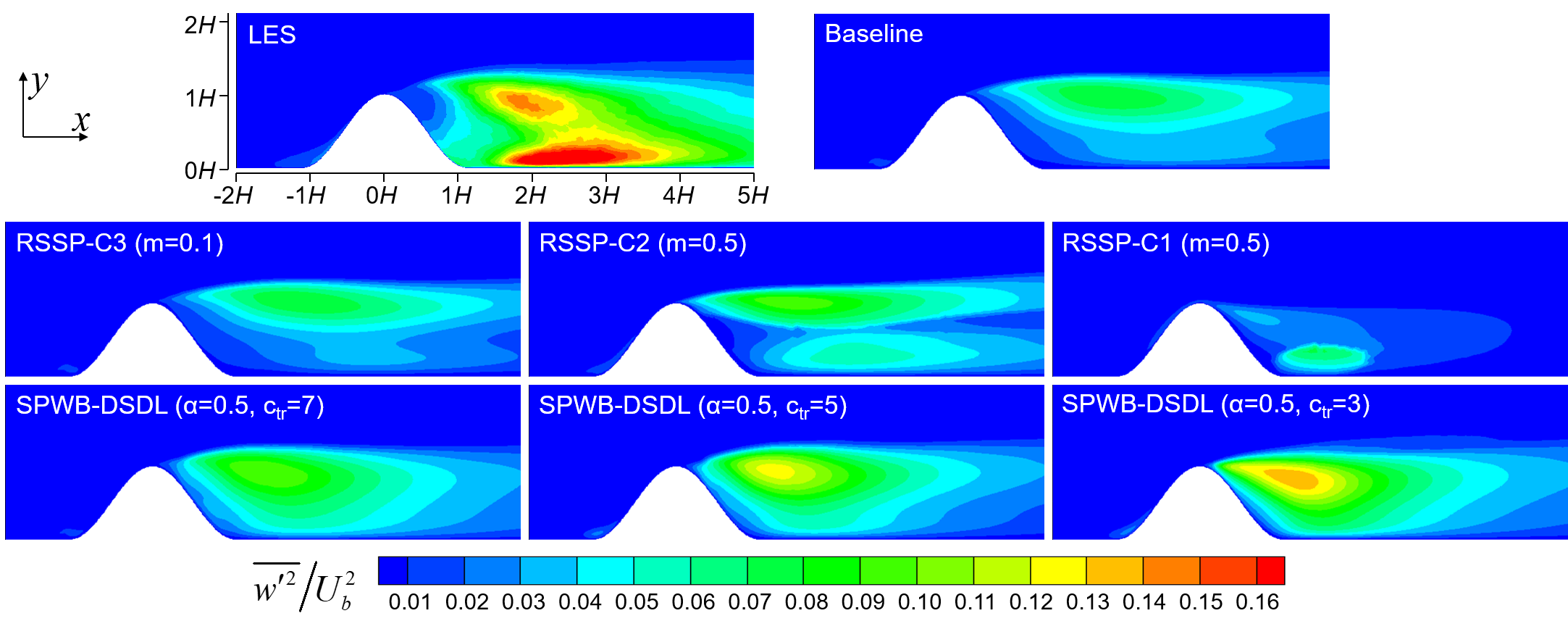}
    \caption{Contours of spanwise turbulence normal stress \(\overline{w^{\prime2}}\) on the plane \(z=0\)}
    \label{fig:Bump_z=0_Rww}
\end{figure}

Compared to the scalar transport in the upwind boundary layer, the scalar transport in the downstream separation bubble exhibits stronger spanwise (\(z\)) variability. Fig.~\ref{fig:Bump_Sherwood2} shows the contours of \(\mathrm{Sh}\) on Patch II. We see that every steady RANS simulation yields one or two \(x\)-directional `stripe' regions with significantly low \(\mathrm{Sh}\), which do not exist in the LES result. These low-\(\mathrm{Sh}\) regions imply the insufficiency of spanwise scalar mixing, a process highly depending on the intensity of spanwise turbulence fluctuations. In Fig.~\ref{fig:Bump_z=0_Rww}, we indeed observe in the LES result a region with significantly large \(\overline{w^{\prime2}}\) in \(y=0\sim0.3H\), while such a region is not predicted by any of the steady RANS simulations. This high-\(\overline{w^{\prime2}}\) region in LES result is associated with the large-scale, quasi-periodic spanwise oscillations of the separation bubble (see \cite{Ching20}). Apparently, modeling these complicated bubble oscillations is beyond the capability of the present DSDL model, which at this stage can only capture the free shear layers as indicated in Figs.~\ref{fig:Bump_z=0_Blending} and \ref{fig:Bump_y=0.5_Blending}. Even so, the SPWB-DSDL model improves upon the baseline model in terms of the average Sherwood number results, \(\mathrm{Sh_{Avg}}\): the error of baseline model is \(-24\%\) while the errors of SPWB-DSDL(\(\beta\)=0.5, \(c_{tr}\)=7.0\(\sim\)3.0) are \(-15\%\sim+2\%\).

\subsubsection{Results of scalar far downstream}
\label{Subsubsec:BumpResultsScalarDownstream}

\begin{figure}[htbp]
    \centering
    \includegraphics[width=1.0\linewidth]{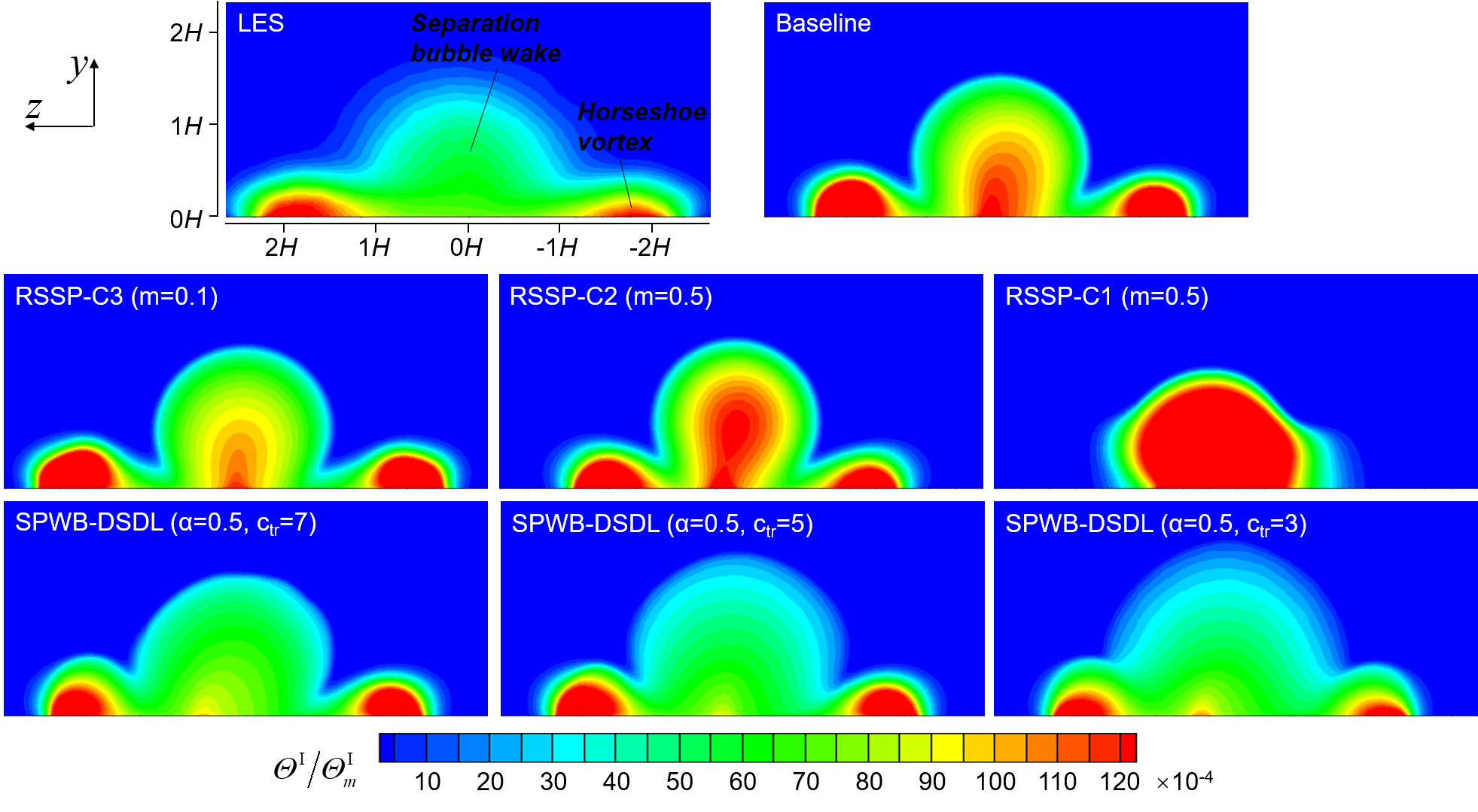}
    \caption{Mean concentration contours of scalar I on the cross section \(x=5H\)}
    \label{fig:Bump_x=5_ScalarUp}
\end{figure}

Lastly, we qualitatively examine the scalar concentrations at a location with a distance downstream from the separation bubble. Fig.~\ref{fig:Bump_x=5_ScalarUp} shows the contours of mean scalar concentration \(\mathit{\Theta}^\mathrm{I}\) on cross-section \(x=5H\). Every result demonstrates that \(\mathit{\Theta}^\mathrm{I}\) is largely confined to regions of the wake induced by the upstream separation bubble and/or of the horseshoe vortex. Compared to the LES result, the baseline and RSSP yield higher \(\mathit{\Theta}^\mathrm{I}\) near the wake center and sharper interface between the wake and mainstream. These discrepancies indicate that the scalar dispersion from the wake to mainstream is underestimated by the baseline and RSSP. This problem is mitigated to some extent by the SPWB-DSDL model, primarily due to its improved \(k\) prediction near the free shear layers between the wake and mainstream.

\begin{figure}[htbp]
    \centering
    \includegraphics[width=1.0\linewidth]{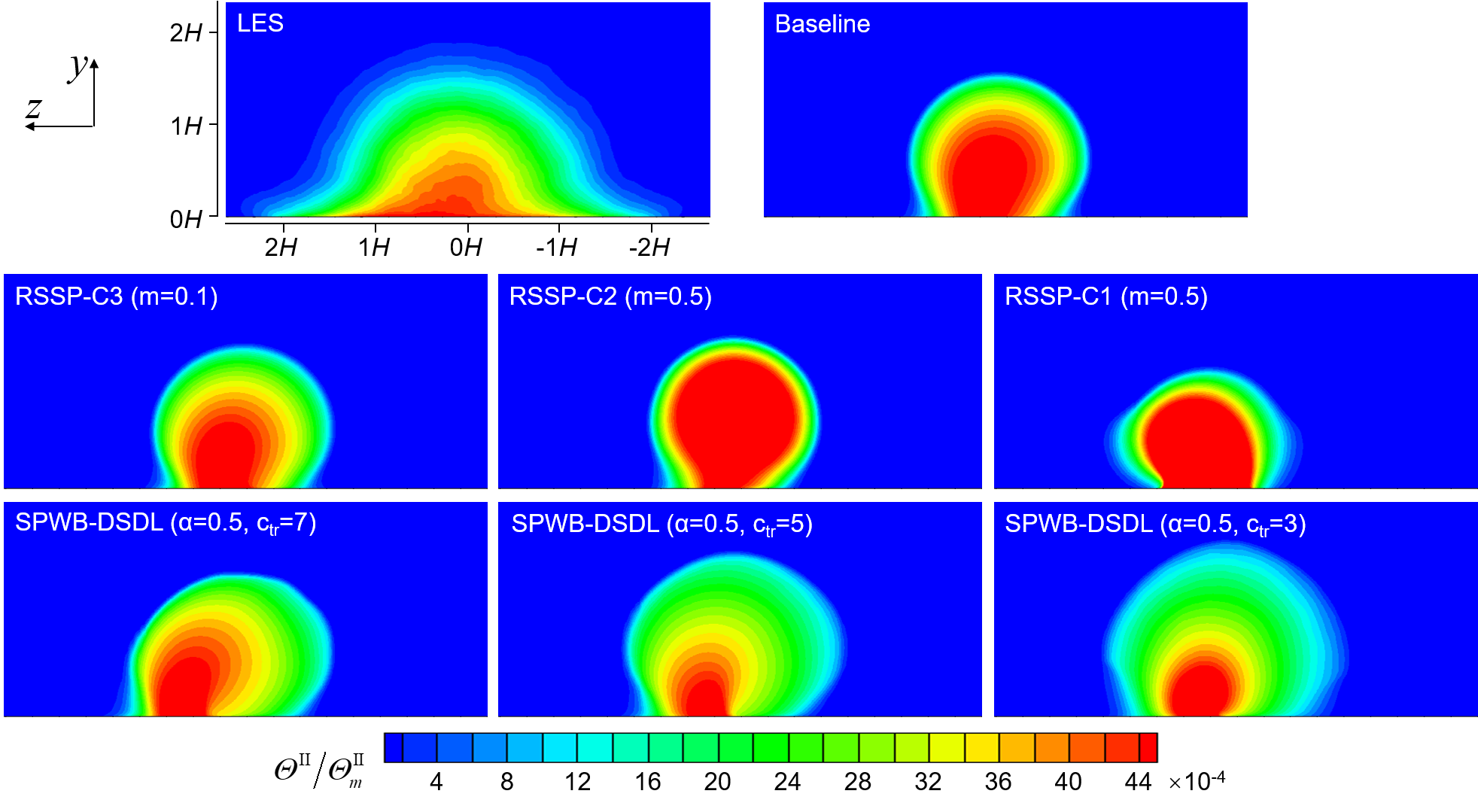}
    \caption{Mean concentration contours of scalar II on the cross section \(x=5H\)}
    \label{fig:Bump_x=5_ScalarDown}
\end{figure}

The above comparative observations on the wake-mainstream dispersion is similarly reflected in Fig.~\ref{fig:Bump_x=5_ScalarDown} that shows the \(\mathit{\Theta}^\mathrm{II}\) contours on the same cross-section. In addition, the LES result in Fig.~\ref{fig:Bump_x=5_ScalarDown} clearly demonstrates the strong spanwise dispersion in the near wall region with \(y=0\sim0.3H\), which is not captured by any steady RANS simulation. This region is consistent with the high-\(\overline{w^{\prime2}}\) region in Fig.~\ref{fig:Bump_z=0_Rww}. It indicates again the importance of the large-scale spanwise oscillations of separation bubble to the scalar dispersion in such a case involving complicated three-dimensional separation.

\subsubsection{Summary}
\label{Subsubsec:BumpSummary}

In this subsection, we first implement an LES for the scalar dispersion around a skewed bump mounted on a wall, thereby building the reference database for this problem. The case features an upstream concave surface flow and a downstream three-dimensional separation around a streamlined obstacle. Second, we implement the steady-RANS simulations using the baseline model, the RSSP method, and the SPWB-DSDL model, and then compare their results against the reference LES data.

Despite the very different configurations, the comparative observations among the steady-RANS results for this bump case are qualitatively similar to those observations for the pin-fin array case in \S\ref{Subsec:PinFinArray}. The baseline EV model overpredicts the size of the separation bubble, and significantly underpredits the turbulence kinetic energy and stress in the free shear layers induced by the separation; it tends to substantially overpredict wall-normal Reynolds stress and scalar transfer on the upwind surface while underpredict the same two properties on the surface covered by the separation bubble; downstream from the bubble, the scalar dispersion from the wake to mainstream is also considerably underestimated. The RSSP method can correct the baseline model's overprediction of bubble size, but is apparently incapable of fixing most of the other above-mentioned problems. By contrast, the SPWB-DSDL model is essentially more capable of fixing most of the above-mentioned problems of the baseline model. In general, among all the steady-RANS simulations considered, the SPWB-DSDL(\(\alpha\)=0.5, \(c_{tr}\)=5.0) realizes the best balance between the predictions of a variety of properties, including bubble size, turbulence energy and stress in bubble, scalar transfer near upwind surface and in bubble, and wake-mainstream scalar dispersion.

One of the most noticeable problems that remain in the SPWB-DSDL model is the lack of a mechanism to capture the large-scale, quasi-periodic spanwise oscillations of the separation bubble, which is crucial to the spanwise scalar mixing in the near-wall region. A mechanism to represent this type of coherent structures could be incorporated into the model in the future.

\section{Conclusions}
\label{Sec:Conclusions}

The present work addresses two common issues relevant to the eddy-viscosity (EV) model failures in steady-RANS simulations for turbulent flows with scalar transport around obstacles. The first issue is the general overprediction of scalar transfer near the upwind surfaces, which is primarily attributed to the absence of wall-blocking mechanism in conventional EV models. We accordingly propose a Shear-Preserving-Wall-Blocking (SPWB) method to analytically correct the overpredicted wall-normal stress while preserving the shear stress under the realizability constraint. The second issue is the general underprediction of scalar transfer in the downstream large separation regions, which is essentially attributed to the presence of vortex shedding invalidating the scaling ground in conventional models' dissipation closures. We accordingly generalize the recently developed Double-Scale Double-Linear-EV (DSDL) model \cite{Hao21} to scalar transport predictions. Consequently, a combined model named SPWB-DSDL is developed. The model includes three free parameters: \(\alpha\sim\mathcal{O}(1)\) controlling the scope of near-wall region in which the wall-blocking correction takes effect; \(c_{tr}\in(1,\infty)\) controlling the intrinsic rate of energy transfer from coherent structures (CS) to stochastic turbulence (ST); and \(\beta\in[0,\infty)\) controlling the CS-ST scale separation effect on the energy transfer.

For comparison, we separately consider a previously developed EV model correction method, termed `Reynolds-stress-shape-perturbation' (RSSP) \cite{Emory13,Gorle12}. Then using the \(k\)-\(\omega\) SST model as the baseline EV model, we investigate the performance of the SPWB and/or DSDL method(s) and the RSSP method in two test cases, respectively. The first case is the forced heat convection in a pin-fin array, featuring upstream impingement flows and downstream quasi-two-dimensional separations around cylindrical obstacles; a previously implemented LES \cite{Hao19} is used as the reference database. The second case is the scalar dispersion around a skewed bump, featuring an upstream concave surface flow and a downstream complicated three-dimensional separation around a streamlined obstacle; an LES is implemented to build the reference database.

The results for both cases indicate the following apparent problems of the baseline EV model in the scenarios of turbulent scalar transport around obstacles:
\renewcommand{\theenumi}{(\arabic{enumi})}
\begin{enumerate}
\item overpredict the mean sizes of the downstream recirculation regions;
\item overpredict the wall-normal Reynolds stress and the scalar transport near the upwind surfaces;
\item underpredict the downstream levels of
turbulence kinetic energy and stress;
\item underpredict the scalar transport between the recirculation or wake regions and the mainstream.
\end{enumerate}
In general, the RSSP method is capable of correcting Problem (1) using C2 or C1 perturbation, but is essentially incapable of correcting Problems (2)\(\sim\)(4) using any of C3, C2 and C1 perturbations. In contrast, the SPWB-DSDL method demonstrates its general capability of effectively correcting Problems (1)\(\sim\)(4) simultaneously. Based on the test cases in this paper, we suggest the free parameter values, which could balance the four corrections to (1)\(\sim\)(4), to be:
\begin{itemize}
\item \(\alpha\approx0.5\);
\item \(c_{tr}\approx1.5\sim1.8\) for bluff obstacles similar to the cylinders, and \(c_{tr}\approx5\) for streamlined obstacles similar to the bump;
\item \(\beta\approx0.5\sim1.0\) for bluff obstacles similar to the cylinders, and \(\beta\approx0\) for strealined obstacles similar to the bump.
\end{itemize}
Further adjustments of these free parameters for more types of obstacles are one of our focuses in the future.

In addition, the results in this work show several remaining problems of the SPWB-DSDL model that are to be solved in the future. First, immediately near the separation onset point, the turbulence kinetic energy and scalar transfer can still be considerably underestimated. To address this problem, we could consider a much stronger CS production formulation while avoiding increasing the shear stress near the separation onset point. Second, for cases involving complicated three-dimensional separations, the model still lacks a mechanism to capture the large-scale spanwise oscillations of
the separation bubble. The future work could consider representing this type of coherent structures in the model.

\bibliographystyle{elsarticle-num-names}
\bibliography{sample.bib}

\end{document}